\documentclass{article}[12pt]
\usepackage{graphicx}
\usepackage{hyperref}
\usepackage{amssymb}
\usepackage{multirow}
\usepackage{caption}
\usepackage{subcaption}
\usepackage{isotope}
\usepackage[left=1.5cm,right=1.5cm,top=1.5cm,bottom=1.5cm]{geometry}
\usepackage{authblk}
\usepackage{amsmath}
\begin{document}
\title{Hybrid compact stars with finite strange quark mass and dark energy: implications for astrophysical observations}
\author[1]{Rohit Roy\thanks{royrohit477@gmail.com}}
\author[2]{Koushik Ballav Goswami\thanks{koushik.kbg@gmail.com}}
\author[3]{Debadri Bhattacharjee\thanks{debadriwork@gmail.com}}
\author[4]{Pradip Kumar Chattopadhyay\thanks{pkc$_{-}76$@rediffmail.com}}
\affil[1,2,3,4]{IUCAA Centre for Astronomy Research and Development (ICARD), Department of Physics, Cooch Behar Panchanan Barma University, Panchanan Nagar, Vivekananda Street, District: Cooch Behar, \\ Pin: 736101, West Bengal, India}
\maketitle
\begin{abstract}{In this work, we present a detailed investigation of compact stars composed of deconfined quark matter, with finite strange quark mass ($m_s\neq0$), admixed with dark energy. The quark sector is described using the MIT bag model equation of state, while the dark energy component is assumed to satisfy a linear equation of state of the form $p^{de}=\omega\rho^{de}$, where $\omega$ lies in the range $-1\leq\omega\leq-\frac{1}{3}$. The stellar configuration is studied within the Finch-Skea ansatz for the $g_{rr}$ metric potential. A coupling between the two components is introduced through the relation $\rho^{de}=\beta\rho^Q$, where $\beta$ denotes the dark energy coupling parameter. From causality arguments, the admissible range of $\beta$ is constrained as $0<\beta<-\frac{1}{3\omega}$. The structural properties of such compact stars are then analysed by varying $\beta$ within this range. Solution of the Tolman-Oppenheimer-Volkoff equations yields a maximum mass of $2.012~ M_{\odot}$ with a corresponding radius of $\approx 11$ km. Notably, for a fixed $\omega$, both the stellar mass and radius are found to decrease with increasing $\beta$. The model is further examined through causality, energy, and stability conditions, all of which are satisfied, thereby ensuring physical viability. Finally, the framework is applied to estimate radii of several compact star candidates identified as strange quark stars admixed with dark energy, with predictions showing close agreement with recent observational results.}
\end{abstract}
\section{Introduction}
\label{sec1}
The final stage of stellar evolution produces a diverse array of compact objects, which include White Dwarfs (WD), Neutron Stars (NS) and Black Holes (BH). Further, NS can be subcategorized as quark stars. The possible existence of ultra dense quark stars was first proposed by many researchers \cite{NT,EF,CA,PH}. As proposed in the Refs. \cite {EF,CAAO,JM} the Strange Quark Matter (SQM), is assumed to be the true ground state of matter made up with almost equal numbers of up ($u$), down ($d$) and strange ($s$) quarks which is also predicted by the SQM hypothesis in the Refs. \cite{EW,HT1,HT2,HT3,HT4}. Distinguishing NS from potential strange stars based on the mass and radius is challenging. It is believed that quark matter may exist in the core of NS \cite{MAPG} or in strange stars, in bulk \cite{JJD}. Recently a wide class of cosmic events such as, the galaxy rotation curve, electromagnetic radiation from type Ia supernova \cite{AGR,SP}, baryon acoustic oscillation and cosmic microwave background radiation \cite{CLB}, has provided high precessional observational data for direct and affirmative evidence of accelerated expansion of the universe. To explain the origin of such accelerated expansion, it is proposed that \cite{TP,RD} a phantom dark energy may be responsible for separating massive bodies one from another more effectively than the attractive gravity force. The entity which is maintaining the expansion of the universe against gravity and kept the universe from collapsing is termed as Cosmological Constant by Einstein. However, the concept of cosmological constant was discarded by Einstein himself. So accelerated nature may be explained if universe may contain some strange kind of physical entity having inherent negative pressure which provides necessary outward force to maintain such expansion of the universe. The expansion rate can be obtained from the Friedman equation with a suitable Equation of State (EoS). While gravitational waves offer a tool for studying dark energy within the framework of General Relativity (GR)\cite{NY1,NY2,BPA}, a recent proposition suggests massive stellar remnants BH, as alternative astrophysical sources for dark energy \cite{DF}.\par 

Chapline \cite{GC} suggested the existence of "dark energy stars", as massive compact objects with quantum critical boundary separating them from ordinary spacetime. The interior of these stars is filled with dark energy. According to Chapline, this boundary acts as a quantum critical shell \cite{GCEH}. When energy exceeds a certain threshold ($Q_0$) \cite{PBTM}, it decays upon entering this shell, releasing radiation perpendicular to the surface. Matter with lower energy can penetrate the shell and travel further into the star. Given that quarks and gluons within nucleons possess energies above this critical threshold $Q_0$ \cite{JB}, dark energy stars in galactic centres could explain the observed excess of positrons through nucleon decay, as predicted by the Georgi-Glashow \cite{HG} grand unified model. Furthermore, these stars might have originated from early universe spacetime fluctuations, analogous to quantum critical instability \cite{GCEH}, offering a potential link to the nature of dark matter. Chapline argued that for a dark energy star any interior fluid with respect to the vacuum Schwarzschild solution may have an EoS of the form $p^{de}=-\frac{1}{3}\rho^{de}$ \cite{GC}. However, E.B Gliner \cite{EBG} first proposed an EoS of the form $p=-\rho$, which latter became a pivotal point for studying dark energy stars. Moreover, the initial decisive solution for that EoS emerged as Bardeen space-time \cite{JMB}. A literature review shows that the EoS parameter $\omega$ is important specifically in the range $-1\leq\omega\leq-\frac{1}{3}$ for stable dark energy stars \cite{PBTM,FSNL,CRG}.\par

Many authors \cite{CRG,AB,PB1,FRRS,GP,SD} have studied the properties of possible dark energy stars. Using continuous dark energy EoS, the structure and stability of compact star in presence of anisotropic pressure has investigated by many researchers \cite{PBTM,AB,PBFR}. At the global scenario the role of quark matter is same as that of dark energy which is an important property as well as an interesting result as predicted by Rahaman et al. \cite{FRPKF} and Brilenkov et al. \cite{MB}. Here, we propose a possible compact star model that consists of quarks having non-zero strange quark mass ($m_s$) admixed with dark energy. Since quark matter may be a source for dark energy \cite{FRPKF,MB}, according to the work of Grammenos et al. \cite{TGR}, we consider a hybrid star model with strange quark matter having finite $m_s$ admixed with dark energy and focus on the effect of $m_s$ in the presence of dark energy to discuss some physical properties that are not well explained using standard models. In our work, the term “dark energy” is used in an effective and local sense, rather than as the homogeneous cosmological component responsible for the accelerated expansion of the Universe. The relation $\rho^{de}=\beta \rho^Q$ should be interpreted as a phenomenological parametrisation of an additional negative-pressure component that emerges under extreme density and strong-gravity conditions. Such an effective contribution may arise from several physically motivated considerations: (i) Vacuum energy modification in dense QCD matter, (ii)  Density-dependent vacuum contributions, (iii) Strong gravity and local energy conditions, (iv) Consistency as an effective model etc.\par 

Moreover, the critical condition under which quark matter is converted into dark energy is yet unknown from standard physics. Some speculative ideas that might relate to the conversion process are given below:
\begin{enumerate}
	\item In extreme conditions (very high densities) inside a compact star, normal nuclear matter may transition into quark–gluon plasma or even more exotic states exhibiting analogical dark energy properties. \cite{GLU}.
	\item Inside a compact star, under extreme pressure, there could be a transition to a lower-energy vacuum state that behaves as a region dominated by vacuum energy, exhibiting  phenomenological dark energy characteristics at small scales. \cite{VAB}.
	\item Quarks may interact with a scalar field associated with dark energy, and in a dense environment this interaction may alter the behaviour of the field, causing it to act as a dark energy source. \cite{GC}. 
	\item Some alternative to BH suggests gravastar, an object where a shell of exotic matter surrounds a core of vacuum rearrangement \cite{POM}. 
	\item Another possibility is that a mechanism which is similar to one that is responsible for the accretion of dark matter could be at work. In this scenario, Weakly Interacting Massive Particles (WIMPs) emerge as potential candidates. These particles could accumulate over time and, because they possess corresponding antiparticles, their mutual annihilation would generate a source of heat. Given the high density expected of dark energy, such heating could significantly influence the internal structure of stars. However, it's important to note that while WIMPs interact weakly with ordinary matter (baryons), particles associated with dark energy are believed to interact only through gravity. As a result, the actual accretion rate would likely be much lower than initially anticipated. 
\end{enumerate}
The proportionality $\rho^de=\beta \rho^Q$ ensures that the additional component remains dynamically relevant in the stellar interior, allowing us to systematically study its impact on equilibrium and stability. We emphasize that our model does not claim that cosmological dark energy is being directly accumulated inside the star. Rather, we employ a dark energy–like term as an effective description of unknown high-density physics or modified vacuum contributions under extreme conditions. Nevertheless, all of these remain a theoretical conjectures. In reality, the mechanism through which dark energy becomes accumulated within stars is not yet understood.\par 
Now, we also consider the hybrid star model to be anisotropic. Anisotropy in compact stars may originate from the presence of superfluidity \cite{RA}, strong magnetic field \cite{FW} or slow rotational motion \cite{HER}. According to Ruderman \cite{Ruderman} highly dense compact objects, where the density surpasses that of nuclear matter, may exhibit anisotropic matter composition. Inside these objects, the existence of type 3A superfluids could contribute to the development of anisotropy within the fluid sphere. Inclusion of the idea of pressure anisotropy in the study of the properties of matter at extremely high densities has also considered by several authors \cite{FRRS,RLB,VV,FR1,FR2,MKALAM,HOSSEIN,MKALAM2,ASAHA}. Bowers and Liang \cite{RLB} emphasised the EoS for relativistic fluid with local anisotropy in pressure and proposed that such anisotropy in pressure may lead to some effects on the various physical features like maximum mass, surface redshift, stiffness of the EoS, radius etc. In the present model, the effective anisotropy arises from the two-fluid description, in which the total radial and tangential pressures are obtained from the combined contributions of the individual fluid components. Thus, the resulting anisotropy is a macroscopic, phenomenological effect of the multi-component stellar configuration and is not introduced through a specific microscopic interaction.
It is worth emphasising that a direct quantitative comparison with anisotropies generated by microscopic mechanisms, such as quark interactions, superfluidity, or strong magnetic fields, is not straightforward because these mechanisms depend on the underlying microphysical equation of state and on additional model-dependent parameters. Nevertheless, the magnitude of the effective anisotropy obtained in our model remains within the range expected for small-to-moderate departures from isotropic pressure support in compact stars. In particular, the anisotropy should be interpreted as an effective description of the internal stress distribution rather than as a direct representation of any one microscopic mechanism.\par 

In the present paper, it is considered that internal fluid consists of a mixture of two type of fluids: $(i)$ quark matter with non-zero strange quark mass ($m_s$) and $(ii)$ dark energy with a repulsive nature. Interaction between these fluids are avoided for the sake of simplicity. To describe the EoS for the hybrid fluid, we use the MIT bag EoS. Various important aspects of compact stars has been extensively analysed by using the MIT bag EoS \cite{AS,PKC,DB,KBG2}. Apart from that, models of quark stars have been studied qualitatively in the Refs. \cite{EA1,EA2,EA3}. While we have a theoretical understanding of potential accumulation of dark matter within stars, due to its gravitational interaction, our current knowledge of physics does not provide a clear mechanism for dark energy to do the same. Dark energy, unlike dark matter, primarily interacts with gravity, making its accretion rate into a star significantly slower. This weaker interaction renders dark energy a negligible contributor to stellar energy production through particle annihilation.\par 

The novelty of the present study lies in extending conventional compact star models by including a dark energy component. We emphasize that this treatment is phenomenological and intended to probe the sensitivity of stellar structure to exotic contributions. The novelty of this study does not lie in proposing an entirely new gravitational framework, but in developing a self-consistent hybrid compact star model that simultaneously incorporates:
\begin{itemize}
	\item finite strange quark mass $(m_s \neq 0)$ within the MIT bag model,
	\item an additional dark energy–like component with negative pressure, and
	\item a proportional coupling between quark matter and dark energy density.
\end{itemize}
While each of these ingredients has been studied separately in the literature, their combined inclusion in a single framework is non-trivial. In particular, the inclusion of finite $m_s$ modifies the equation of state in a non-linear way, affecting pressure and energy density. The dark energy coupling parameter ($\beta$) introduces an additional degree of freedom that alters the equilibrium configuration. The interplay between $m_s$ and $\beta$ leads to measurable deviations in macroscopic observables such as mass–radius relation, compactness, and stability bounds. Our contribution is an extension toward a more realistic and coupled description, rather than a fundamentally new theory. The approach builds on established frameworks (Einstein field equations, MIT bag EoS, Finch–Skea geometry). However, the work is not merely parameter fitting. Specifically, the Einstein field equations are solved for a two-fluid anisotropic system, which is structurally different from standard single-fluid models. The proportionality relation $\rho^{de} = \beta \rho^{Q}$ introduces a new effective coupling, leading to modified Tolman-Oppenheimer-Volkoff (TOV) equations. The presence of finite strange quark mass explicitly alters thermodynamic quantities, which propagates through the full solution. Thus, the model represents a modified physical system with coupled components, rather than a simple optimisation of existing equations. It is clarified that the present work is phenomenological in nature, intended to explore the possible role of exotic components under extreme conditions. The “dark energy” considered here is not the cosmological dark energy, but an effective negative-pressure contribution that may arise from unknown high-density physics or vacuum modifications. Given the current lack of a fundamental theory describing such effects at compact star densities, this approach serves as an exploratory framework. To address this concern, we have: $(i)$ explicitly stated the effective nature of the dark energy component, and $(ii)$ discussed possible physical motivations, and emphasised that the model aims to provide testable deviations in observable quantities (mass, radius, stability).\par 

In this article, we investigate the effects in various physical parameters of a star due to the inclusion of non-zero strange quark mass ($m_s$) and dark energy. Following the previous works of \cite{CRG,PB1,PR}, it is assumed that linear proportionality may exist between the densities of dark energy and quarks. To find the various physical parameters we solve the Einstein field equations (EFE) in the Finch-Skea geometry. Now, using the most general form of the EoS, we solve the TOV equation to find the maximum mass and radius of the hybrid star in the presence of dark energy. For stable strange matter relative to $\isotope[56]{\it Fe}$, the range of the bag constant ($B_g$) is taken as $57.55~MeV/fm^3<B_g<95.11~MeV/fm^3$, which, however, is modified because of the presence of a finite value of $m_s(\neq0)$ \cite{RR,KBG}. For the analysis we carefully choose $B_g$ such that the quark system remains in the stable region, i.e., the energy per baryon lies below $930.4~MeV$ at zero external pressure condition, which is the value of binding energy per baryon of $\isotope[56]{\it Fe}$ \cite{JM}. However, according to Bodmer \cite{ARB} and Witten hypothesis \cite{EW} inside the star, non-zero pressure may push the binding energy per baryon ($E_B$) above $930.4~MeV$ and SQM may be stable above these value of $E_B$. While our model builds upon established frameworks such as the MIT bag model, the Finch–Skea metric, and a linear dark energy equation of state, the novelty of our work lies in the combined and self-consistent incorporation of the finite strange quark mass $(m_s)$ along with dark energy effects. In much of the existing literature on quark–dark energy hybrid stars, simplifying assumptions such as massless quarks $(m_s=0)$ are often adopted. However, the inclusion of a finite $m_s$ introduces non-trivial modifications to the equation of state, affecting the pressure, energy density, and chemical equilibrium conditions. This, in turn, has a direct impact on the mass–radius relation, compactness, and stability criteria of the stellar configuration. Furthermore, the interplay between strange quark mass, dark energy coupling (through $\beta$) leads to distinguishable quantitative deviations in the macroscopic properties of the star compared to earlier models. Therefore, the present work is an extension that captures a more realistic physical scenario by incorporating $m_s$ alongside dark energy within a unified framework. Further observational discriminants would strengthen the model, and this will be an important direction for future investigation.\par 

The outline of this paper is as follows: Sec.~\ref{sec2} delves into the thermodynamics of quarks considering a non-zero strange quark mass. In Sec.~\ref{sec3}, the Einstein field equations has been solved to determine the metric potential and establish constraints on various physical parameters. The maximum value for mass, radius, and central density are presented in Sec.~\ref{sec4}. A comprehensive analysis of the model is provided in Sec.~\ref{sec5}. Sec.~\ref{sec6} examines different stability conditions. Finally, Sec.~\ref{sec7} summarises the key findings of our study. 

\section{Thermodynamics of quarks}
\label{sec2}
To determine the EoS for quark matter, we must study the thermodynamics of quarks. It is hypothesised that at sufficiently high densities and pressure, neutrons are compressed to form deconfined quarks, which remain in $\beta$-equilibrium phase and maintain overall charge neutrality condition. These quarks form a colour singlet baryon having baryon number $A$. As each neutron is made up of three quarks and quarks being fermions, the deconfined quark assembly can be approximated as a Fermi gas of $3A$ quarks. In the MIT bag model \cite{CHK,AC}, the quark confinement can be described by the dynamical equations given below:

\begin{equation}
	p^Q=\sum_{i=u,d,s,e^{-}}p_i-B_g,\label{1}
\end{equation}
\begin{equation}
	\rho^Q=\sum_{i=u,d,s,e^{-}}\rho_i+B_g.\label{2}
\end{equation}
where $\rho_i$ represents the energy density of the $i^{th}$ type particle and $p_i$ is the corresponding pressure. The superscript $Q$ stands for quark. The interior matter of a star is assumed to be composed mainly of three types of quarks up, down and strange along with electrons ($e^{-}$). Muons are completely abundant below the threshold density for charm quarks (if present). Additionally, the possibility of presence of charm quarks is not possible, as charm stars are not stable against the radial oscillations \cite {CHK}. Therefore, the overall charge neutrality condition of such star is given by the following condition:
\begin{equation}
	\sum_{i=u,d,s,e^{-}}  n_iq_i=0,\label{3}
\end{equation}
where $q_i$ is the charge of $i^{th}$ particle and $n_i$ is the number density of $i^{th}$ type particle, respectively. Expressions for $p_i$, $\rho_i$ and $n_i$ can be evaluated from the thermodynamic potential:
\begin{equation}
	d\Omega_i=-S_idT-P_idV-N_id\mu_i.\label{4}
\end{equation}
From Eq.~(\ref{4}), one can obtain the following expressions:
\begin{equation}
	p_i=\frac{g_i}{6\pi^2}\int_{m_i}^{\infty}\left(E_i^2-m_i^2\right)^{3/2}f(E_i)\,dE_i,\label{5}
\end{equation}
\begin{equation}
	\rho_i=\frac{g_i}{2\pi^2}\int_{m_i}^{\infty}E_i^2\left(E_i^2-m_i^2\right)^{1/2}f(E_i)\,dE_i,\label{6}
\end{equation}
\begin{equation}
	n_i=\frac{g_i}{2\pi^2}\int_{m_i}^{\infty}E_i\left(E_i^2-m_i^2\right)^{1/2}f(E_i)\,dE_i,\label{7}
\end{equation}
where, $E_i^2=k^2+m_i^2$ and $m_i$ is the mass of the $i^{th}$ particle. The term $g_i$ is known as degeneracy factor. $g_i=6$ for quarks of all colours and for leptons is $g_i=2$. $f(E_i) =\frac{1}{1+e^{-(E_i-\mu_i)/T}}$ is the Fermi-Dirac distribution function. The chemical equilibrium between the different quark flavours as well as electrons is established through the weak interactions as follows:
\begin{equation}
	d\rightarrow u+e^{-}+\bar{\nu}_{e^-},\label{8}
\end{equation}
\begin{equation}
	s\rightarrow u+e^{-}+\bar{\nu}_{e^-},\label{9}
\end{equation}
\begin{equation}
	s+u \longleftrightarrow d+u. \label{10}
\end{equation}
As neutrinos continuously escape from the stellar interior, their chemical potential can be equated to zero. Hence, to maintain necessary condition of equilibrium, we must have,
\begin{equation}
	\mu_d=\mu_u+\mu_e,\label{11}
\end{equation}
\begin{equation}
	\mu_s=\mu_u+\mu_e,\label{12}
\end{equation}
and 
\begin{equation}
	\mu_s=\mu_d\equiv\mu. \label{13}
\end{equation}
where $\mu_{j}$ represents chemical potential of $j^{th}$ particle ($j=u,d,s,e^{-}$). Since the temperature of the star is much lower than the chemical potentials of quarks given in eqs.~(\ref{11})-(\ref{13}), we can consider approximately that the temperature of the star to be close to zero ($T\rightarrow 0$). According to work of Kettner et al. \cite{CHK} it is considered that $u$ and $d$ quarks have negligible mass. However, we consider the strange quark has a non-zero mass ($m_s$). In our model, $m_s=0$ does not imply the absence of strange quarks; rather, it corresponds to the limiting case where the strange quark is treated as massless, similar to up and down quarks. Thus, following the expressions from eqs.~(\ref{5})-(\ref{7}), we have:
\begin{equation}
	p_i=\frac{g_i\mu^4\eta_i^4}{24\pi^2}\left[\sqrt{1-z_i^2}(1-\frac{5}{2}z_i^2)+\frac{3}{2}z_i^4ln\frac{1+\sqrt{1-z_i^2}}{z_i}\right],\label{14}
\end{equation}
\begin{equation}
	\rho_i=\frac{g_i\mu^4\eta_i^4}{8\pi^2}\left[\sqrt{1-z_i^2}(1-\frac{1}{2}z_i^2)-\frac{1}{2}z_i^4ln\frac{1+\sqrt{1-z_i^2}}{z_i}\right],\label{15}
\end{equation}
\begin{equation}
	n_i=\frac{g_i\mu^3\eta_i^3}{8\pi^2}\left[1-z_i^2\right]^{\frac{3}{2}},\label{16}
\end{equation}
where, $z_i$ is defined as given in Ref.~\cite{CHK}:
\begin{equation}
	z_i=\frac{m_i}{\mu_i}\label{17}
\end{equation}
In this context overall charge neutrality condition defined in eq.~(\ref{3}) can be represented by the following necessary condition:
\begin{equation}
	2\left(1-\frac{\mu_{e}}{\mu}\right)^3-\left(\frac{\mu_{e}}{\mu}\right)^3-\left[1-\left(\frac{m_s}{\mu}\right)^2\right]^{3/2}-1=0\label{18}
\end{equation}
Using eqs.~(\ref{14})-(\ref{17}), we obtain a relationship between the radial pressure ($p_r^Q$) and energy density ($\rho^Q$) of quark matter for a finite value of mass of strange quark:
\begin{equation}
	p_r^Q=\frac{1}{3}\Big(\rho^Q-4B_1\Big).\label{19}
\end{equation}
where $B_{1}=\frac{4B_g+\rho_s-3p_s}{4}$, $\rho_s$ and $p_s$ represent the energy density and pressure for the strange quark respectively. In the expression of $B_1$, $B_g$ is bag constant. On the other hand, $\rho_s$ and $p_s$ depend on the strange-quark mass as well as chemical potential and, consequently, on the local thermodynamic state. Therefore, unlike $B_g$, $B_1$ is not a constant independent of the stellar matter. Rather, it may be termed as an effective bag parameter defined through the effect of strange-quark mass on the thermodynamic potential.

\section{Solution of the Einstein field equation}
\label{sec3}
The line element associated with interior space-time of a static and spherically symmetric fluid sphere is represented as:
\begin{equation}
	ds^2=-e^{\nu(r)}dt^2+e^{\lambda(r)}dr^2+r^2(d{\theta^2}+\sin^2{\theta}\;d{\phi^2}),\label{20}
\end{equation}
where $e^{\nu(r)}$ and $e^{\lambda(r)}$ are the $g_{tt}$ and $g_{rr}$ metric components to be determined and depend only on the radial coordinate $r$. In GR the connection between matter and geometry of space-time is represented by the following equation:
\begin{equation}
	R_{\mu\nu}-\frac{1}{2}g_{\mu\nu}R=\frac{8{\pi}G}{c^4}T_{\mu\nu},\label{21}
\end{equation}
where, $R_{\mu\nu}$ and $R$ are termed as Ricci tensor and Ricci scalar, respectively, and $G$ represents Newtonian gravitational constant. Now, as the internal matter is composed of quarks in the presence of dark energy, the energy-momentum tensor $T_{\mu\nu}$ in mixed tensor form for such a mixture is given as;
\begin{eqnarray}
	T^{\mu}_{\nu}=(p_r^{total}-p_t^{total})\chi^{\mu}\chi_{\nu}+(\rho^{total}+p_t^{total})v^{\mu}v_{\nu}-p_t^{total}\delta^{\mu}_{\nu} =\mbox{diag}~(-\rho^{total},p_r^{total},p_t^{total},p_t^{total})
	,\label{22} 
\end{eqnarray}
where, $v^{\mu}v_{\mu}=-\chi^{\mu}\chi_{\mu}=1$, and $v^{\mu}\chi_{\mu}=0$, the vector $v^{\mu}$ is the fluid four-velocity and $\chi^{\mu}$ is the unit spacelike vector in the radial direction which is orthogonal to $v^{\mu}$. It can be verified that $\nabla^{\nu} T_{\mu\nu}=0$ indicating conservation of energy-momentum in GR. $\rho^{total}=\rho^Q+\rho^{de}$ is the total energy density, $p_r^{total}=p_r^Q+p_r^{de}$ and $p_t^{total}=p_t^Q+p_t^{de}$ represent the total values of pressure at radial and tangential directions respectively. Here, the superscripts 'Q' and 'de' represent the quark and dark profiles of energy density and pressure, respectively.  Combining eqs.~(\ref{20}), (\ref{21}) and (\ref{22}), the following equations has been obtained (taking $8{\pi}G=1,\;c=1$):
\begin{equation}
	\rho^{total}=\frac{\lambda^{\prime}e^{-\lambda}}{r}+\frac{1-e^{-\lambda}}{r^2},\label{23}
\end{equation}
\begin{equation}
	p_r^{total}=\frac{\nu^{\prime}e^{-\lambda}}{r}-\frac{1-e^{-\lambda}}{r^2},\label{24}
\end{equation}
\begin{equation}
	p_t^{total}=e^{-\lambda}\left[\frac{\nu^{\prime\prime}}{2}+\frac{{\nu^{\prime}}^2}{4}+\frac{\nu^{\prime}-\lambda^{\prime}}{2r}-\frac{\nu^{\prime}\lambda^{\prime}}{4}\right].\label{25}
\end{equation}
where, overhead dash represents derivative with respect to $r$. Here we considered that $\Delta=(p_t^{total}-p_r^{total})$ is the measure of pressure anisotropy as proposed in the references \cite{SDM,MKM,KD,MC}. To solve the above set of equations we need to specify an EoS for the dark energy component. In the literatures \cite{TGR,CRG,KANIKA,FR}, it is shown that the density of dark energy is simply proportional to the density of quark energy, as follows:
\begin{equation}
	\rho^{de}=\beta\rho^Q.\label{26}
\end{equation}
where, $\beta(>0)$ is a proportionality constant known as coupling parameter that represents the fraction of quark energy which is converted and presented as dark energy. The relation $\rho^de=\beta \rho^Q$ is introduced as a phenomenological ansatz aimed at exploring possible signatures of dark energy within ultra-dense compact objects. To date there is no well-established theoretical framework that describes the microphysical coupling between dark energy and quark matter at stellar scales. Our aim is to parametrize a potential effective contribution that may arise from unknown high-density physics or modified gravity effects. It should be emphasized that we do not assume that cosmological dark energy literally accumulates inside the star; rather, it is treated as an effective fluid component representing possible vacuum or repulsive effects that may arise at ultra-high densities. Regarding the accumulation of dark energy inside compact stars, we assume that under extreme gravitational fields, dark energy may not remain entirely homogeneous and could exhibit clustering or effective coupling with matter degrees of freedom. Similar phenomenological approaches have been adopted in the literature to probe such possibilities. The parameter $\beta$ thus serves as a measure of the strength of this coupling, allowing us to systematically study its impact on the equation of state and macroscopic observables. Notably, it models a proportional coupling between the dark energy density and the quark (or quark–gluon) energy density, suggesting that these components evolve together through a common physical mechanism or interaction. Such a relation can arise if dark energy is influenced by the quantum vacuum or condensate effects of quark fields, as in models where vacuum energy depends on the quark condensate or the QCD trace anomaly. $\beta$ then encodes the strength of this coupling. In essence, this assumption provides a simplified phenomenological link between dark energy and the quark sector, allowing the study of possible interactions or feedback between QCD-scale physics and cosmic acceleration.\par  
Additionally, following the references \cite{AB,CRG,BD}, the dark energy EoS can be represented as, 
\begin{equation}
	p_r^{de}=\omega\rho^{de}.\label{27}
\end{equation}
In presence of dark energy with repulsive pressure, we choose the range of $\omega$ to be $-1\leq\omega\leq-\frac{1}{3}$ \cite{PBTM,BD}. According to the Friedmann equation, $\omega<-\frac{1}{3}$ represents cosmic expansion. On the other hand $\omega=-1$ corresponds to the special case of cold dark matter with cosmological constant \cite{AB}. Combining eq.~(\ref{19}) and eq.~(\ref{27}), we have an expression for the total radial pressure for the system as:
\begin{equation}
	p_r^{total}=\frac{\left(\frac{1}{3}+\omega\beta\right)}{1+\beta}\rho^{total}-\frac{4B_1}{3}.\label{28}
\end{equation}  
Eq.~(\ref{28}) is the effective EoS of matter composed of $3$-flavour quarks with non-zero value of strange quark mass ($m_s$) coupled with dark energy. Now, to solve eqs.~(\ref{23})-(\ref{25}), we choose the $g_{rr}$ component of the metric to be Finch-Skea \cite{FS} type as follows:
\begin{equation}
	e^{\lambda}=1+ar^2.\label{29}
\end{equation}
where, $a$ is an unknown constant to be determined. In the present work, the use of the Finch–Skea ansatz is intended to provide an analytical framework that allows us to obtain closed-form solutions and gain qualitative insight into the role of model parameters (such as $m_s$, $\beta$, and $\omega$) on the internal structure. It serves as a complementary approach rather than a replacement for the standard TOV integration. Using eqs.~(\ref{23}), (\ref{26}) and (\ref{29}), we have the following expressions for the quark and total energy density:
\begin{equation}
	\rho^Q=\frac{a(3+ar^2)}{(1+ar^2)^2(1+\beta)},\label{30}
\end{equation}
\begin{equation}
	\rho^{total}=(1+\beta)\rho^Q.\label{31}
\end{equation}
Now from Eqs.~(\ref{24}), (\ref{28}) and (\ref{30}), we can solve for the $g_{tt}$ component of the metric which is expressed as,
\begin{equation}
	\nu(r)=\frac{1}{3(1+\beta)}\left[\frac{1}{2}(1+ar^2)(4+3\beta+3\beta\omega)-\frac{B_1(1+ar^2)^2(1+\beta)}{a}+(1+3\beta\omega)\log (1+ar^2)\right]+c,\label{32}
\end{equation}   
where, $c$ is an integration constant. Now, the expressions for total radial and tangential pressure can be obtained from Eqs.~(\ref{24}), (\ref{25}) which are given as follows:
\begin{eqnarray}
	p_r^{total}=\frac{1}{3(1+ar^2)^2(1+\beta)}\Bigg[a^2r^2\left(1+4B_1r^2(1+\beta)+3\beta\omega\right)+a\left(3-8B_1r^2(1+\beta)+9\beta\omega\right)\nonumber\\-4B_1(1+\beta)\Bigg],\label{33}
\end{eqnarray}
\begin{eqnarray}
	p_t^{total}=\frac{1}{36(1+ar^2)^3(1+\beta)^2}\Bigg[16B_1(1+\beta)^2(B_1r^2-3)a^4r^6\Biggl\{4-4B_1r^2(1+\beta)+3\beta(1+\omega)\Biggr\}^2\nonumber\\+4a(1+\beta)\Biggl\{9+16B_1^2r^4(1+\beta)+27\beta\omega-6B_1r^2\Big(9+8\beta+3\beta\omega\Big)\Biggr\}+2a^3r^4\Biggl\{32B_1^2r^4(1+\beta)^2\nonumber\\+3\Big(3+2\beta+3\beta\omega\Big)\Big(4+3\beta(1+\omega)\Big)-4B_1r^2(1+\beta)\Big(23+3\beta(6+5\omega)\Big)\Biggr\}+a^2r^2 \Biggl\{60+96B_1^2r^4\nonumber\\(1+\beta)^2-8B_1r^2(1+\beta)\left(40+33\beta+21\beta\omega\right)+3\beta \Big(26+42\omega+3\beta\left(3+\omega(8+9\omega)\right)\Big)\Biggr\}\Bigg].\label{34}
\end{eqnarray}
From Eqs.~(\ref{33}) and (\ref{34}) the value of total anisotropy $\Delta$ can be obtained through the relation as $\Delta=\Big(p_t^{total}-p_r^{total}\Big)$.

\subsection{Bounds on model parameters}
\label{sec3.1}
For our proposed model to be realistic and physically viable, the following bounds are imposed on various model parameters:
\begin{itemize}
	\item The total energy density at the centre of the star is $\rho^{total}_c=3a$. This implies that constant $a$ should be positive in nature.
	\item In the interior of a compact star with non-zero $m_s$ coupled with dark energy, the radial ($v_r$) and tangential ($v_t$) components of the sound velocity must satisfy the causality condition. The condition represents that the square of the radial ($v_{r}^2$) and tangential velocity ($v_{t}^2$) of sound must lie within the range $0\leq v_{r}^2 \leq 1$, $0\leq v_{t}^2 \leq 1$ or equivalently $0\leq(\frac{dp_r^{total}}{d\rho^{total}})\leq 1$ and $0\leq(\frac{dp_t^{total}}{d\rho^{total}})\leq 1$. Now from eqs.~(\ref{31}) and (\ref{33}), we get the expression for $v_r^2$ is as given below:
	\begin{equation}
		v_r^2=\frac{1+3\omega\beta}{3(1+\beta)}.\label{35}
	\end{equation} 
	To satisfy the causality condition, the above expression puts a restriction on the parameter $\beta$ which is given as $-1<\beta\leq-\frac{1}{3\omega}$. However, negative value of $\beta$ is not physically allowed in view of (\ref{26}). Therefore for a physically viable model of a star with dark energy profile, the value of $\beta$ should lie within the range $0<\beta\leq-\frac{1}{3\omega}$.
	\item At the stellar surface the total radial pressure should vanish, i.e., $p_r^{total}(r=R)=0$ \cite{mafa}. 
	\item To find the unknowns $a$ and $c$, the solutions of interior and exterior regions should be matched at the stellar surface ($r=R$). Now, the exterior Schwarzschild solution is given below:
	\begin{equation}
		ds^2=-\left(1-\frac{2M}{r}\right)dt^2+\left(1-\frac{2M}{r}\right)^{-1}dr^2  +r^2\left(d{\theta^2}+\sin^2{\theta}\;d{\phi^2}\right).\label{36}
	\end{equation}
	where $M$ is the total mass. Using the matching condition we obtain,
	\begin{equation}
		e^{\nu(r=R)}=e^{-\lambda(r=R)}=(1-\frac{2M}{R}).\label{37}
	\end{equation}
\end{itemize}

\begin{figure}
	\centering
	\begin{minipage}{0.48\textwidth}
		\centering
		\includegraphics[width=\linewidth]{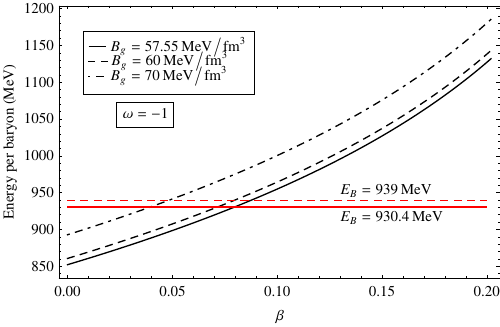}
		\subcaption{}\label{fig1}
	\end{minipage}
	\hfil
	\begin{minipage}{0.48\textwidth}
		\centering
		\includegraphics[width=\linewidth,height=0.65\linewidth]{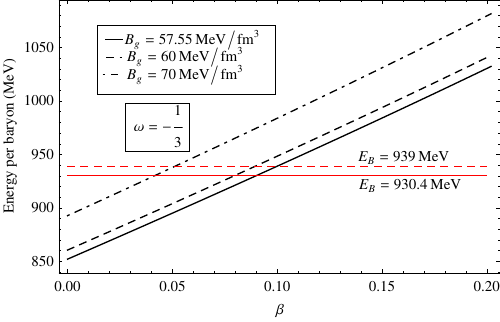}
		\subcaption{}\label{fig2}
	\end{minipage}
	\caption{Panel (a) Variation of $E_B$ with coupling parameter $\beta$ for $m_s=100~MeV$ and $\omega=-1$. The solid, dashed and dotdashed lines represent different bag value as $B_g=57.55~MeV/fm^3$, $60~MeV/fm^3$ and $70~MeV/fm^3$. Panle (b) Variation of $E_B$ with coupling parameter $\beta$ for $m_s=100~MeV$ and $\omega=-\frac{1}{3}$. The solid, dashed and dotdashed lines represent different bag value as $B_g=57.55~MeV/fm^3$, $60~MeV/fm^3$ and $70~MeV/fm^3$.}
\end{figure}

Now, to solve for the chemical potentials ($\mu$ and $\mu_e$) and energy per baryon ($E_B$) the conditions that total pressure, $p_r^{total}=0$ at the surface along with charge neutrality condition as given in eq.~(\ref{18}) has been implemented. The energy per baryon ($E_B$) is plotted as a function of dark energy coupling parameter $\beta$, for parametric choices of $B_g$ and $m_s=100~MeV$. The respective variations of $E_B$ are shown in Figs.~\ref{fig1} and \ref{fig2} for $\omega=-1$ and $-\frac{1}{3}$, respectively. In the context of bulk quark matter, it is standard to express thermodynamic quantities in terms of the baryon number density ($n_B=\frac{1}{3} n_q$), which is a conserved and physically meaningful quantity in compact star studies. Accordingly, the energy per baryon represents the energy per unit baryon number of the system, providing a convenient normalization to compare the stability of quark matter with hadronic matter. This quantity is widely used to assess whether strange quark matter is energetically favourable (e.g., in comparison to nuclear matter benchmarks). Notably, $E_B$ increases with $\beta$ and this increment is higher if $B_g$ is increased. This may be explained as follows: the total energy density consists of $\rho^Q$ and $\rho^{de}$. Now, increasing $\beta$ results in the increase of $\rho^{total}$. On the other hand, the baryon number density depends only on $\rho^Q$, as $\rho^Q$ decreases with increase in $\beta$, this leads to the increase of $E_B$. Again, depending on the value of $E_B$, a stable region ($E_B<930.4~MeV$) can be ensured below which the $3$-flavour quark matter is energetically stable with respect to zero pressure condition \cite{JM}. Also a probable metastable region ($930.4~MeV<E_B<939~MeV$) and an unstable region ($E_B>939~MeV$) can also be defined depending on the value of $E_B$. Thus depending on the stability window one may have some limits on the value of coupling parameter $\beta$, which are listed in Table~\ref{tab1} and it is evident that bag constant ($B_g$) has some effects on the dark energy coupling parameter $\beta$. 
\begin{table}[ht]
	\centering
	\caption{Table for range of $\beta$ depending on value of $E_B$ at zero pressure condition, taking $m_s=100~MeV$.}\label{tab1}
	\resizebox{0.6\textheight}{0.065\textwidth}{$
		\begin{tabular}{@{}c|ccc|ccc}
			\hline
			$B_g$ & \multicolumn{3}{c}{$\omega=-1$} \vline& \multicolumn{3}{c}{$\omega=-\frac{1}{3}$}  \\ \cline{2-7}
			$MeV/fm^3$ & stable region & Meta-stable region & unstable region & stable region & Meta-stable region & unstable region \\ \hline
			57.55     & $\beta\leq 0.080$ & $0.080<\beta\leq 0.088$ & $\beta>0.088$ & $\beta\leq 0.090$ & $0.090<\beta\leq 0.100$ & $\beta>0.100$ \\
			60         & $\beta\leq 0.072$ & $0.072<\beta\leq 0.08$ & $\beta>0.080$ & $\beta\leq 0.080$ & $0.080<\beta\leq 0.089$ & $\beta>0.890$ \\
			70         & $\beta\leq 0.040$ & $0.040<\beta\leq 0.048$ & $\beta>0.048$ & $\beta\leq 0.042$ & $0.042<\beta\leq 0.051$ & $\beta>0.051$ \\ \hline
		\end{tabular}$}
\end{table}
depending on which nature of SQM is found to be stable, metastable or unstable. If the value of $B_g$ increases, $\beta$ picks up lower value as tabulated in Table~\ref{tab1}. It also has been noted that for $B_g>80~MeV/fm^3$, the SQM does not attend stability even if $\beta$ is as low as $0.001$. 

\section{Maximum mass}
\label{sec4}
\begin{figure}
	\centering
	\begin{minipage}{0.48\textwidth}
		\centering
		\includegraphics[width=\linewidth]{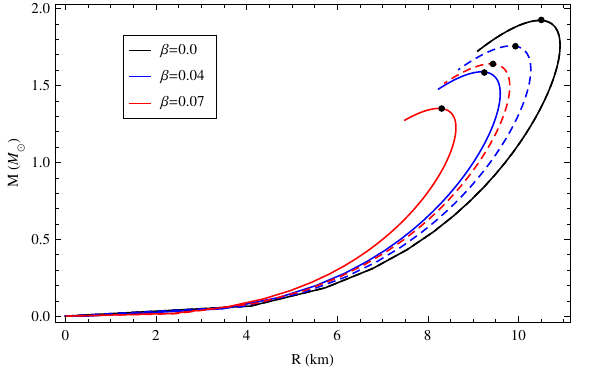}
		\subcaption{}\label{fig3}
	\end{minipage}\hfil
	\begin{minipage}{0.45\textwidth}
		\centering
		\includegraphics[width=\linewidth,height=0.65\linewidth]{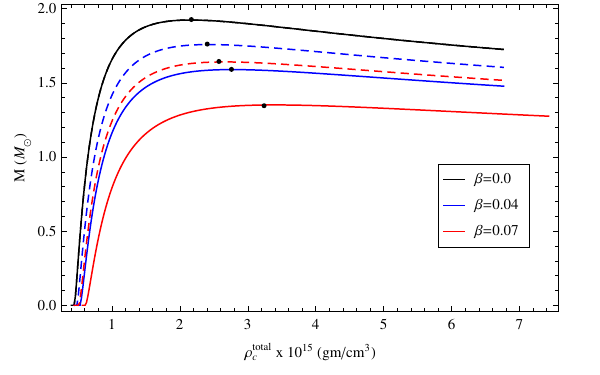}
		\subcaption{}\label{fig4}
	\end{minipage}\hfil
	\caption{Panel (a) Mass-radius variation for different value of $\beta$ for $m_s=100~MeV$ and $B_g=60~MeV/fm^3$. The solid and dashed lines represent $\omega=-1$ and $-\frac{1}{3}$ respectively. The black dots represent the maximum mass value for each curve. Panel (b) Plot of maximum mass with total central density ($\rho^{total}$) for different value of $\beta$ with $m_s=100~MeV$ and $B_g=60~MeV/fm^3$. The solid and dashed lines represent $\omega=-1$ and $-\frac{1}{3}$ respectively. }
\end{figure}

\begin{figure}
	\centering
	\includegraphics[width=8cm]{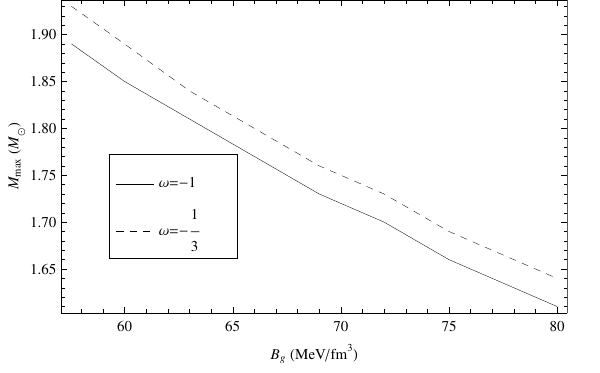}
	\caption{Variation of maximum mass with bag parameter ($B_g$). Here solid and dashed lines represent $\omega=-1$ and $-\frac{1}{3}$ respectively.}\label{fig4a}
\end{figure}

\begin{figure}
	\centering
	\begin{minipage}{0.45\textwidth}
		\includegraphics[width=0.85\linewidth,height=0.6\textwidth]{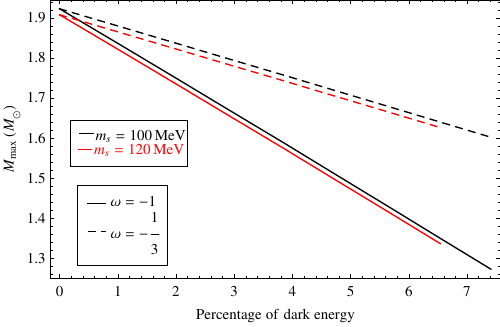}
		\subcaption{}\label{fig5}
	\end{minipage}\hfil
	\begin{minipage}{0.45\textwidth}
		\centering
		\includegraphics[width=0.85\linewidth,height=0.6\textwidth]{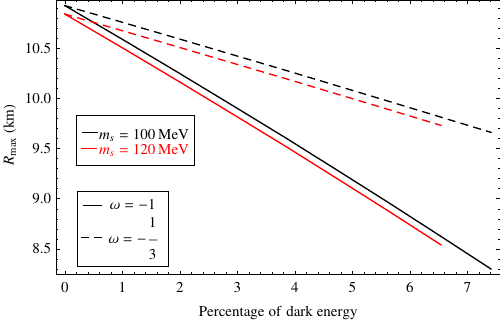}
		\subcaption{}\label{fig6}
	\end{minipage}\hfil
	\caption{Panel(a) Variation of maximum mass with the percentage of dark energy for $B_g=60~MeV/fm^3$. Here, the black and red lines represent the corresponding variation for a parametric choice of $m_s=100~MeV$ and $120~MeV$, respectively. Also the solid and dashed lines represent the choice of $\omega=-1$ and $\omega=-\frac{1}{3}$, respectively. Panel (b) Variation of maximum radius with percentage of dark energy for $B_g=60~MeV/fm^3$. Here, the black and red lines represent the corresponding variation for a parametric choice of $m_s=100~MeV$ and $120~MeV$, respectively. Also the solid and dashed lines represent the choice of $\omega=-1$ and $\omega=-\frac{1}{3}$, respectively. }
\end{figure}

The active gravitational mass contained within a sphere of radius $r$ depends on the total energy density ($\rho^{total}$) and must increase with the increase of the radius \cite{NKG} and is maximal at a particular value of $r=R$ (radius of the star). Now, the mass function, i.e., the gravitational mass is obtained from the following expression:
\begin{equation}
	m(r)=4\pi\int_{0}^{r}\rho^{total}\;r^2dr.\label{38}
\end{equation}
It is evident from Eq.~(\ref{38}) that $m(r)=0$ for $r=0$ and that $m(r)=M$, when $r=R$, $M$ being the total mass contained within the radius $R$. The compactness factor is given as $u(R)=\frac{M}{R}$. Again for a spherically symmetric perfect fluid in static equilibrium, the total mass and radius should satisfy the Buchdahl criterion, $\frac{M}{R}\leq\frac{4}{9}$ \cite{BUDH}. Now, to solve for maximum mass we need to solve the coupled TOV equations which are given as:  
\begin{equation}
	\frac{dm}{dr}=4\pi r^2\rho^{total},\label{39} 
\end{equation}
and 
\begin{equation}
	\frac{dp_r^{total}}{dr}=-(\rho^{total}+p_r^{total})\Big[\frac{m}{r^2}+4\pi rp_r^{total}\Big]\left(1-\frac{2m}{r}\right)^{-1}.\label{40}
\end{equation}
Solving TOV equations with the EoS as given in eq.~(\ref{28}), the mass-radius relation for different fractions of dark energy ($\beta$) are shown in Fig.~\ref{fig3} for a parametric choice of $m_s=100~MeV$ and $B_g=60~MeV/fm^3$. The black dots on each curve represent values of the maximum mass and corresponding radius. The maximum mass and radius are tabulated in Table~\ref{tab2} for different parametric $m_s$ and $\beta$ value. It is interesting to note that with the increase of $\beta$ for a fixed $\omega$ and $m_s$, reduces the maximum mass ($M_{max}$). This may be attributed to the fact that if the percentage of dark energy ($\frac{\beta}{1+\beta}\times100$) increases, it necessarily replaces some fraction of the dense quark matter. Although this conversion/phase change increases $\rho^{total}$, it decreases $p_r^{total}$ and as a result the EoS becomes softer. While dark energy provides a repulsive (negative pressure) contribution, its effect on stellar structure is not solely to increase the mass. In our model, increasing the coupling parameter $\beta$ enhances the dark energy component, which effectively reduces the net attractive gravitational binding and softens the overall equation of state. As a result, the star cannot sustain as much mass against gravitational equilibrium, leading to a decrease in the maximum mass. Thus, although dark energy is repulsive, its dominance weakens the self-gravitating system, limiting the mass that can be stably supported. Hence, $M_{max}$ gets reduced. Again for a fixed $\beta$ and $\omega$, increasing $m_s$ results in a lower value of pressure for a given energy density of the compact star, which also refers to a softer EoS indicating that the compact object can be further compressed and as a result maximum mass and radius picks up smaller values. Table~\ref{tab2} also emphasizes that the decrease in $M_{max}$ and $R_{max}$ with the increase of $\beta$ are more prominent for $\omega=-1$ than for $\omega=-\frac{1}{3}$. However, the total central density increases with increasing $\beta$ value. In Fig.~\ref{fig4}, variation of mass with central density are shown for different $\beta$ value and it is observed that the Harrison-Zeldovich-Novikov stability criterion \cite{YBZ,BKH} is fulfilled for the data set used here up to the maximum central density ($\rho_c^{total}$). In Fig.~\ref{fig4a}, we have shown graphically the dependence of maximum mass ($M_{max}$) on bag parameter ($B_g$). It can be noted that the maximum mass decreases monotonically with bag parameter ($B_g$). In Figs.~\ref{fig5} and \ref{fig6}, the variations of maximum value of mass and radius with dark energy percentage ($\frac{\beta}{1+\beta}\times100$) are shown for $\omega=-1$ and $\omega=-\frac{1}{3}$, respectively, along with a parametric choice of $m_s$ and $B_g$.\par

The most significant physical parameters of a star are the energy density ($\rho$), pressure ($p$), surface redshift ($Z_s$), total mass ($M$), radius ($R$) as well as moment of inertia ($I$) in case of slow and rigid rotation. While, most variables are not directly observable, the surface redshift ($Z_s$) is a measurable parameter. When a photon is moving radially outward from the surface of a star, it change its frequency as seen by an observer and is measured by the quantity surface redshift $Z_s$. The surface redshift is a function of $M$ and $R$ and its expression is given below \cite{GHB}:
\begin{equation}
	Z_s=\left(1-\frac{2M}{R}\right)^{-1/2}-1.\label{41}
\end{equation} 
Now form eq.~(\ref{41}), $Z_s$ seems to be the maximum for the maximum compactness ($\frac{M}{R}$). For a physically realistic isotropic star, $Z_s$ must follow the limit $Z_s\leq{2}$ \cite{BUDH,NS,CGB}, whereas the limit is modified owing to the introduction of pressure anisotropy by Ivanov \cite{Ivanov} and is expressed as $Z_s<5.211$ \cite{Ivanov}. 
\begin{table}[ht]
	\centering
	\caption{Table for maximum mass ($M_{max}$), radius ($R_{max}$) and total central density ($\rho_c^{total}$) for $B_g=60~MeV/fm^3$.}\label{tab2}
	\resizebox{0.62\textwidth}{!}{$
		\begin{tabular}{@{}c|c|cccc}
			\hline
			$\omega$ & $m_s$  & $\beta$ & $M_{max}$ & $R_{max}$ & $\rho_c^{total}\times{10^{15}}$ \\
			& ($MeV$)  &         &  ($M_\odot)$ &   ($km$)       &  ($gm~cm^{-3}$)                \\  \hline
			\multirow{9}{*}{-1} & \multirow{3}{*}{80} & 0.0 & 1.94 & 11.01 & 1.96  \\ 
			&                     & 0.04 & 1.60 & 9.68  & 2.37 \\
			&                     & 0.06 & 1.44 & 9.03 & 2.56 \\ \cline{2-6} 
			& \multirow{3}{*}{100}& 0.0 & 1.92 & 10.92 & 1.96  \\ 
			&                     & 0.04 & 1.59 & 9.60  & 2.36 \\
			&                     & 0.06 & 1.43 & 8.95 & 2.67 \\ \cline{2-6}
			& \multirow{3}{*}{120}& 0.0 & 1.91 & 10.84 & 1.96  \\ 
			&                     & 0.04 & 1.57 & 9.52  & 2.39 \\
			&                     & 0.06 & 1.41 & 8.86 & 2.79 \\ \hline 
			\multirow{9}{*}{$-\frac{1}{3}$} & \multirow{3}{*}{80} & 0.0 & 1.94 & 11.01 & 1.96  \\ 
			&                     & 0.04 & 1.77 & 10.36  & 2.13 \\
			&                     & 0.06 & 1.69 & 10.05 & 2.21 \\ \cline{2-6} 
			& \multirow{3}{*}{100}& 0.0 & 1.92 & 10.92 & 1.96  \\ 
			&                     & 0.04 & 1.76 & 10.28  & 2.15 \\
			&                     & 0.06 & 1.68 & 9.96 & 2.30 \\ \cline{2-6}
			& \multirow{3}{*}{120}& 0.0 & 1.91 & 10.84 & 1.96  \\ 
			&                     & 0.04 & 1.74 & 10.19 & 2.22 \\
			&                     & 0.06 & 1.66 & 9.88 & 2.36 \\ \hline                
		\end{tabular}$}	
\end{table} 

\section{Physical viability of the present model}
\label{sec5}
In this article, we explore the physical features of hybrid stars with finite strange quark mass ($m_s$) mixed with dark energy and determine the necessary constraints on the parameters so that the solutions are found to be physically viable. We also perform some studies on a large number of recently observed compact objects which are assumed to be such hybrid stars. In Table~\ref{tab3}, we have shown predicted radii of such compact objects for their observed mass and using the condition $p_{r}^{total}(R)=0$, where $r=R$ defines the surface of the compact object. It is worthwhile to note that the theoretically obtained radius agrees well with observational results.\par
\begin{figure}
	\centering
	\begin{minipage}{0.45\textwidth}
		\centering
		\includegraphics[width=0.85\linewidth,height=0.6\textwidth]{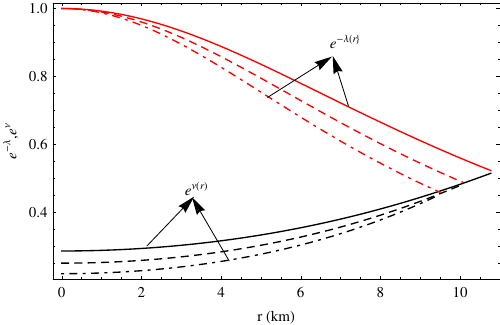}
		\subcaption{}\label{fig7}
	\end{minipage}\hfil
	\begin{minipage}{0.45\textwidth}
		\centering
		\includegraphics[width=0.85\linewidth,height=0.6\textwidth]{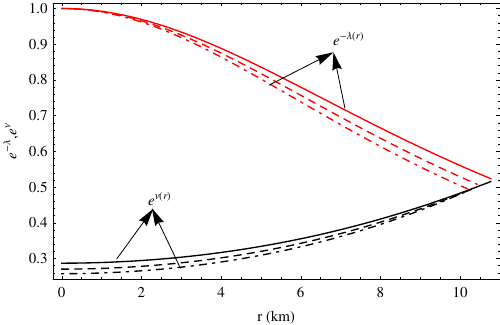}
		\subcaption{}\label{fig8}
	\end{minipage}\hfil
	\caption{Panel (a) Variation of different metric potentials inside $4U~1608-52$ taking $m_s=100~MeV$, $B_g=60~MeV/fm^3$ and $\omega=-1$. The solid, dashed and dotdashed lines represent $\beta=0$, $0.04$ and $0.07$ respectively. Panel (b) Variation of different metric potentials inside $4U~1608-52$ taking $m_s=100~MeV$, $B_g=60~MeV/fm^3$ and $\omega=-\frac{1}{3}$. The solid, dashed and dotdashed lines represent $\beta=0$, $0.04$ and $0.07$ respectively.}
\end{figure}

	
For a detailed study, we select a compact object namely $4U~1608-52$ having the mass $1.74~M_{\odot}$ and radius of $9.3\pm1~km$ \cite{Guver}. To study the effect of $m_s$ and $\beta$ on physical parameters of the star, such as metric potentials, total energy density, total radial $(p_r^{total})$ and transverse pressure $(p_t^{total})$ along with quark profiles and total anisotropy, graphical analysis has been employed. The variations of $e^{-\lambda}$ and $e^{\nu}$ are shown in Figs.~\ref{fig7} and \ref{fig8} for $\omega=-1$ and $\omega=-\frac{1}{3}$, respectively, with a parametric choice of $m_s=100~MeV$, $B_g=60~MeV/fm^3$ and different $\beta$ values. From Figs.~\ref{fig7} and \ref{fig8}, it is interesting to note that with the increase of $\beta$, the value of $e^{-\lambda}$ at all interior points of the star decreases while at the centre of the star $\beta$ has no effect on $e^{-\lambda}$ for both cases $\omega=-1$ and $\omega=-\frac{1}{3}$. On the other hand, at the interior of the star $e^{\nu}$ decrease while $\beta$ increases, for $\omega=-1$. For $\omega=-\frac{1}{3}$, same type of variation of $e^{\nu}$ is obtained at all internal points. At the surface, $\beta$ has no effect on $e^{\nu}$.

\begin{figure}
	\centering
	\begin{minipage}{0.45\textwidth}
		\centering
		\includegraphics[width=0.85\linewidth,height=0.6\textwidth]{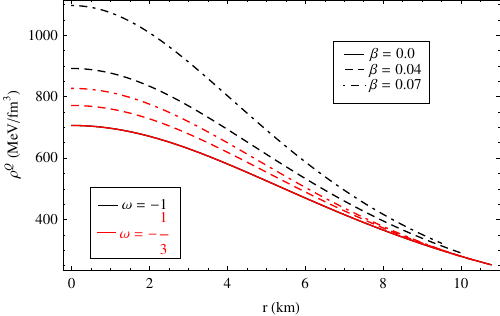}
		\subcaption{}\label{fig9}
	\end{minipage}\hfil
	\begin{minipage}{0.45\textwidth}
		\centering
		\includegraphics[width=0.85\linewidth,height=0.6\textwidth]{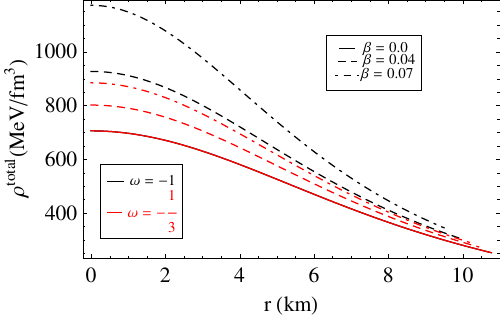}
		\subcaption{}\label{fig10}
	\end{minipage}
	\caption{Panel (a) Radial variation of $\rho^Q$ inside $4U~1608-52$ with $m_s=100~MeV$, $B_g=60~MeV/fm^3$ for $\omega=-1$ (black line) and $\omega=-\frac{1}{3}$ (red line). The solid, dashed and dotdashed lines represent $\beta=0$, $0.04$ and $0.07$ respectively. Panel (b) Radial variation of $\rho^{total}$ inside $4U~1608-52$ taking $m_s=100~MeV$, $B_g=60~MeV/fm^3$ for $\omega=-1$ (black line) and $\omega=-\frac{1}{3}$ (red line). The solid, dashed and dotdashed lines represent $\beta=0$, $0.04$ and $0.1$ respectively.}
\end{figure}

\begin{figure}
	\centering
	\includegraphics[width=0.55\textwidth]{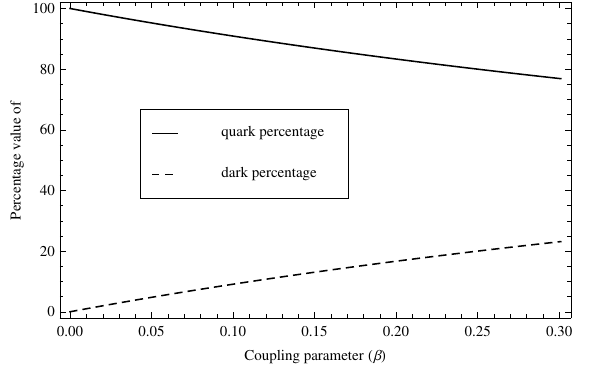}
	\caption{Variation of quark and dark percentage of energy with coupling parameter $\beta$.}\label{fig11}
\end{figure}

\begin{figure}
	\centering
	\begin{minipage}{0.45\textwidth}
		\centering
		\includegraphics[width=0.85\linewidth,height=0.6\textwidth]{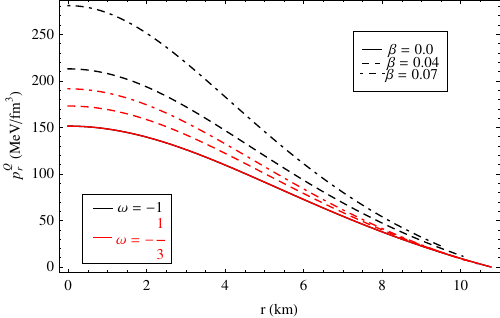}
		\subcaption{}\label{fig12}
	\end{minipage}\hfil
	\begin{minipage}{0.45\textwidth}
		\centering
		\includegraphics[width=0.85\linewidth,height=0.6\textwidth]{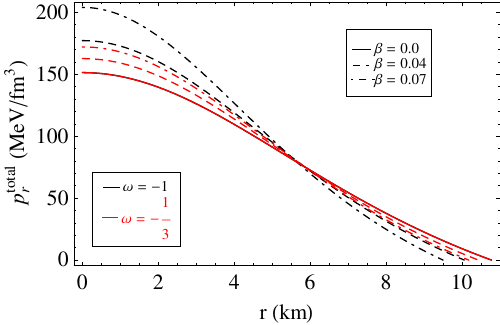}
		\subcaption{}\label{fig13}
	\end{minipage}
	\caption{Panel (a) Radial variation of $p_r^Q$ inside $4U~1608-52$ taking $m_s=100~MeV$, $B_g=60~MeV/fm^3$ for $\omega=-1$ (black line) and $\omega=-\frac{1}{3}$ (red line). The solid, dashed and dotdashed lines represent $\beta=0$, $0.04$ and $0.07$ respectively. Panel (b) Variation of $p_r^{total}$ inside $4U~1608-52$ with $r$ taking $m_s=100~MeV$, $B_g=60~MeV/fm^3$ for $\omega=-1$ (black line) and $\omega=-\frac{1}{3}$ (red line). The solid, dashed and dotdashed lines represent $\beta=0$, $0.04$ and $0.07$ respectively.}
\end{figure}

\begin{figure}
	\centering
	\includegraphics{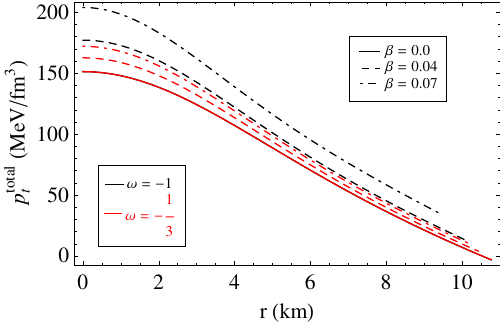}
	\caption{Radial variation of $p_t^{total}$ inside $4U~1608-52$ for $m_s=100~MeV$, $B_g=60~MeV/fm^3$ for $\omega=-1$ (black line) and $\omega=-\frac{1}{3}$ (red line). The solid, dashed and dotdashed lines represent $\beta=0$, $0.04$ and $0.07$ respectively.}\label{fig14}
\end{figure}

\begin{figure}
	\centering
	\begin{minipage}{0.45\textwidth}
		\centering
		\includegraphics[width=0.85\linewidth,height=0.6\textwidth]{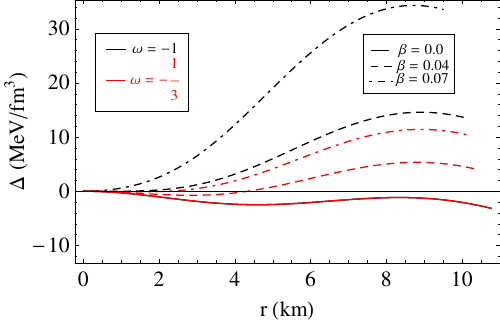}
		\subcaption{}\label{fig15}
	\end{minipage}\hfil
	\begin{minipage}{0.45\textwidth}
		\centering
		\includegraphics[width=0.85\linewidth,height=0.6\textwidth]{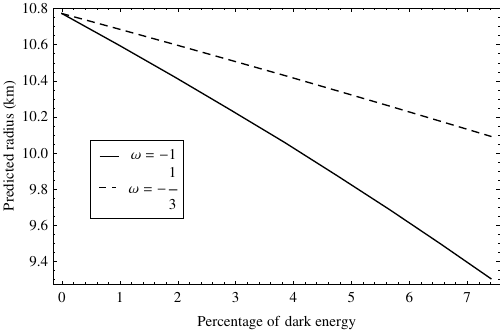}
		\subcaption{}\label{fig16}
	\end{minipage}
	\caption{Panel (a) Variation of $\Delta$ with respect to $r$ for different dark energy percentages inside $4U~1608-52$ for a parametric choice of $m_s=100~MeV$, $B_g=60~Mev/fm^3$ for $\omega=-1$ (black line) and $\omega=-\frac{1}{3}$ (red line). The solid, dashed and dotdashed lines represent $\beta=0$, $0.04$ and $0.07$ respectively. Panel (b) Variation of the predicted radius of $4U~1608-52$ with the percentage of dark energy for $B_g=60~MeV/fm^3$ and $m_s=100~MeV$. The solid and dashed lines represent the variation for $\omega=-1$ and $\omega=-\frac{1}{3}$, respectively. }
\end{figure}

Since the metric potentials are associated with the mass and radius of a star, both $m_s$ and $\beta$ have distinct effects on the stellar mass and the associated radius. Figs.~\ref{fig9} and \ref{fig10} show that the central value of energy density of quarks ($\rho^Q$) and total energy density ($\rho^{total}$) are positive and decrease monotonically towards the stellar surface. Notably, from Figs.~\ref{fig9} and \ref{fig10}, it is evident that the central values of quark energy density ($\rho^Q$) and also the total energy density ($\rho^{total}$) increases with the increment of $\beta$ for both the cases of $\omega=-1$ and $-\frac{1}{3}$. As $\rho^{de}$ is proportional to $\rho^{Q}$ as per eq.~(\ref{26}), therefore, with the increase of $\beta$, $\rho^{de}$ also increases. Again, the increment of $\rho^{de}$ is $\beta$ times the increment of $\rho^Q$. Hence, the contribution of $\rho^{de}$ towards the increment in $\rho^{total}$ is higher than that of $\rho^Q$. This also indicate that the percentage of $\rho^{de}$ ($=\frac{\rho^{de}}{\rho^{total}}$) increases with $\beta$, however, the percentage of $\rho^Q$ ($=\frac{\rho^Q}{\rho^{total}}$) decreases with $\beta$, which is as shown in Fig.~\ref{fig11}. The variation of radial pressure profiles $p_r^Q$ and $p_r^{total}$ are shown in Figs.~\ref{fig12} and \ref{fig13}, with suitable parametric choice of $m_s$, $B_g$ and $\beta$. Variations of the total tangential pressure is shown in Fig.~\ref{fig14} for a parametric choice of $m_s=100~MeV$ and $B_g=60~MeV/fm^3$ with different $\beta$. The variation of total anisotropy ($\Delta$) is shown in Fig.~\ref{fig15}. Fig.~\ref{fig15} shows that, although $\Delta$ is zero at the centre irrespective of value of $\beta$, $\Delta$ has a negative nature for $\beta=0$ for both the choice of $\omega=-1$ and $\omega=-\frac{1}{3}$. With an increase in $\beta$ value $\Delta$ increases and reaches its maximum value near the surface. These types of anisotropic profiles are found in the work of Mak and Harko \cite{MH}. In Fig.~\ref{fig16}, the variation of the predicted radius with the dark energy percentage for $4U~1608-52$ is shown using different parametric choices of $B_g$ and $m_s$. As observed in Fig.~\ref{fig16}, the predicted radius decreases with increasing dark energy percentage. In Table~\ref{tab4}, values of the total central energy density ($\rho_c^{total}$), total surface density ($\rho_0^{total}$), total central radial pressure ($p_{rc}^{total}$), mass to radius ratio at surface ($u=\frac{M(R)}{R}$) and surface redshift ($Z_s$) are tabulated. Notably, the value of $u$ and $Z_s$ satisfy their respective limits.  

\begin{table}[ht]
	\centering
	\caption{Predicted radius for different compact objects with finite strange quark mass ($m_s=100~MeV$) mixed with dark energy for $B_g=60~MeV/fm^3$.}\label{tab3}
	\resizebox{0.8\textwidth}{!}{$
		\begin{tabular}{@{}c|c|c|ccc|ccc}
			\hline
			Name of Compact  & Observed      & Measured      & \multicolumn{3}{c}{Predicted radius for $\omega=-1$} \vline & \multicolumn{3}{c}{Predicted radius for $\omega=-\frac{1}{3}$}\\ \cline{4-9}
			stars            &  mass         &   radius        & $\beta$ & Percentage of    & $R$    & $\beta$ & Percentage of   & Radius \\
			&  $(M_\odot)$  &   $(km)$        &         & dark energy      & $(km)$ &         & dark energy  & $(km)$ \\ \hline
			$Vela~X-1$       & 1.77\cite{TG}  & 9.56$\pm$0.08  & 0.068   & 6.37             & 9.56   & 0.0801  & 7.41         & 10.13 \\
			$4U~1608-52$     & 1.74\cite{Guver}&9.3$\pm$1      & 0.080   & 7.41             & 9.30   & 0.0800  & 7.40         & 10.09 \\
			$Cen~X-3$        & 1.49\cite{TG}  & 9.178$\pm$0.13 & 0.072   & 6.72             & 9.17   & 0.0800  & 7.40         & 9.78 \\
			$HER~X-1$        & 0.85\cite{TG}  & 8.1$\pm$0.41   & 0.060   & 5.66             & 8.17   & 0.0800  & 7.40         & 8.48 \\ 
			$PSR~J-1903+327$ & 1.667\cite{TG} & 9.438$\pm$0.03 & 0.069   & 6.45             & 9.44   & 0.0801  & 7.41         & 10.01 \\ 
			$4U~1820-30$     & 1.58\cite{TG}  & 9.1$\pm$0.4    & 0.078   & 7.24             & 9.18   & 0.0801  & 7.41         & 9.84 \\ 
			$HESS-1731-347$  & 0.77\cite{PC}  & 10.4$\pm$0.86  & 0.050   & 4.76             & 10.11  & 0.0600  & 5.66         & 10.65\\ 
			$PSR~J0030+451$  & 1.34\cite{TER} & 12.71$\pm$1.14 & 0.055   & 5.21             & 11.18  & 0.0680  & 6.37         & 11.94\\ \hline 
		\end{tabular}$}	
\end{table}

\begin{table}[ht]
	\centering
	\caption{Table for total energy density at the centre ($\rho_c^{total}$), total energy density at surface ($\rho_0^{total}$), total central radial pressure ($p^{total}_{rc}$), compactness ($u=\frac{M}{R}$) and surface redshift ($Z_s$) for the value of the parameters listed in Table~\ref{tab3} and taking $\omega=-1$.}\label{tab4}
	\resizebox{0.8\textwidth}{!}{$
		\begin{tabular}{@{}c|ccccc}
			\hline
			Name of the    &$\rho_c^{total}\times10^{15}$ & $\rho_0^{total}\times10^{14}$ & $p^{total}_{rc}\times10^{35}$ & u & $z_s$\\
			compact star   & $gm~cm^{-3}$             & $gm~cm^{-3}$              & $dyne~cm^{-2}$        &     & \\ \hline
			$Vela~X-1$     & 2.10                    & 6.08                    & 3.35                  & 0.2730 & 0.4844  \\
			$4U~1608-52$   & 2.27                    & 6.45                    & 3.44                  & 0.2758 & 0.4935  \\
			$Cen~X-3$      & 1.75                    & 6.19                    & 2.47                  & 0.2395 & 0.3854  \\
			$HER~X-1$      & 1.06                    & 5.84                    & 1.10                  & 0.1533 & 0.2010  \\ 
			$PSR~J-1903+327$&1.95                    & 6.10                    & 2.98                  & 0.2604 & 0.4447  \\
			$4U~1820-30$   &1.96                     & 6.38                    & 2.82                  & 0.2539 & 0.4255  \\ 
			$HESS~1731-347$&0.73                     & 4.44                    & 0.74                  & 0.1293 & 0.1614  \\
			$PSR~J0030+451$&0.98                     & 4.44                    & 1.48                  & 0.1941 & 0.2785  \\ \hline 
		\end{tabular}$}
\end{table}

\begin{figure}
	\centering
	
\end{figure}
\newpage
\subsection{Energy Condition}
\label{sec5.1}
The energy conditions associated with the fluid coupled with dark energy is studied in this section. In GR, energy conditions are applied to determine the physically acceptable energy momentum tensor associated with the fluid distribution. Such energy conditions, in general, are very helpful in determining the inherent properties of such fluid distribution. It is interesting to note that in such analysis information on exact specification of the structure of internal matter is not required. Therefore, without requisite knowledge on energy density of composed matter or its pressure, physical characteristic associated with some astrophysical phenomena, such as the collapse of gravitating body or presence of space-time singularity may be evaluated. In the context of relativistic astrophysics, the analysis of necessary energy conditions, in principle, is an algebraic problem \cite{CAK}, more precisely the problem of eigenvalues associated with the energy momentum tensor. In particular, if fluid violates the strong, weak, null and dominant energy conditions, the general relativity can not be taken into consideration. In four dimensional framework, the study of such energy conditions related to a fluid corresponds to solutions of polynomial equation of degree four to obtain its roots. It is very much complicated due to complexity in obtaining solutions in analytical form for eigenvalues. However, a fluid of physically realistic nature must follow the above mentioned energy conditions simultaneously, the derivation of general solutions for the roots are difficult \cite{CAK,SWH,RW}. In the present model energy conditions can mathematically be expressed as \cite{BP,BPB}:

\begin{enumerate}
	\item DEC: $\rho^{total}\geq 0,~\rho^{total}\;-p_{r}^{total}\geq 0,~\rho^{total}\;-p_{t}\geq 0$.
	\item WEC: $\rho^{total}+p_{r}^{total}\geq 0,~\rho^{total}\geq{0},\;\rho^{total}+p_{t}^{total}\geq{0}$.
	\item NEC: $\rho^{total}+p_{r}^{total}\geq{0},~\rho^{total}+p_{t}^{total}\geq{0}$.
	\item SEC: $\rho^{total}+p_{r}^{total}\geq 0,~\rho^{total}+ p_{r}^{total}+ 2p_{t}^{total}\geq 0$. 
\end{enumerate}

\begin{figure}
	\centering
	\begin{minipage}{0.45\textwidth}
		\centering
		\includegraphics[width=0.85\linewidth,height=0.55\textwidth]{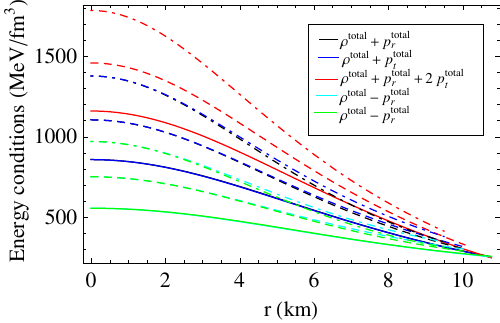}
		\subcaption{}\label{fig17}
	\end{minipage}\hfil
	\begin{minipage}{0.45\textwidth}
		\centering
		\includegraphics[width=0.85\linewidth,height=0.55\textwidth]{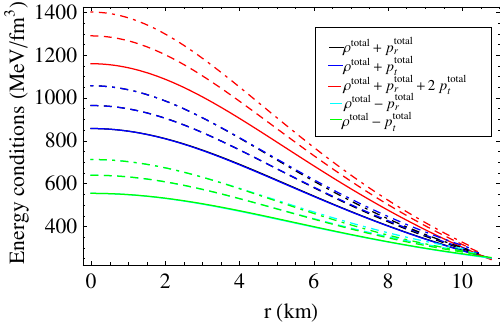}
		\subcaption{}\label{fig18}
	\end{minipage}\hfil
	\caption{Panel (a) represents variation of different energy conditions with $r$ inside $4U~1608-52$ for $B_g=60~MeV/fm^3$, $m_s=100~MeV$ and $\omega=-1$ and in Panel (b) for $m_s=100~MeV$ and $\omega=-\frac{1}{3}$. Here the solid, dashed and dotdashed lines represent the values of $\beta$, respectively, as $0.0$, $0.04$ and $0.07$. The black, blue, red, cyan and green lines represent $\rho^{total}+p_r^{total}$, $\rho^{total}+p_t^{total}$, $\rho^{total}+p_r^{total}+2p_t^{total}$, $\rho^{total}-p_r^{total}$ and $\rho^{total}-p_t^{total}$, respectively.}
\end{figure}

The variation of different energy conditions inside $4U~1608-52$ are shown in Figs.~\ref{fig17} and \ref{fig18} for a parametric choice of model parameters. From Figs.~\ref{fig17} and \ref{fig18}, it is observed that all the required energy conditions are satisfied within the parameter space used here.

\subsection{Causality condition}
\label{sec5.2}
\begin{figure}
	\centering
	\begin{minipage}{0.45\textwidth}
		\centering
		\includegraphics[width=0.85\linewidth,height=0.55\textwidth]{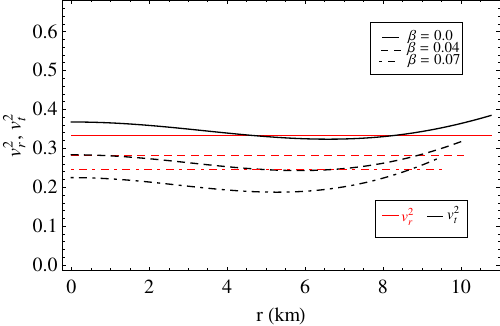}
		\subcaption{}\label{fig19}
	\end{minipage}\hfil
	\begin{minipage}{0.45\textwidth}
		\centering
		\includegraphics[width=0.85\linewidth,height=0.55\textwidth]{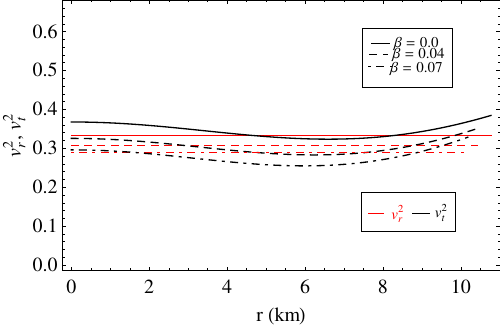}
		\subcaption{}\label{fig20}
	\end{minipage}\hfil
	\caption{Radial variation of $v_r^2$ and $v_t^2$ taking $m_s=100~MeV$, $B_g=60~MeV/fm^3$ and $\omega=-1$. The solid, dashed and dot-dashed lines represent $\beta=0$, $0.04$ and $0.07$ respectively in Panel (a) and in Panel (b) for $m_s=100~MeV$, $B_g=60~MeV/fm^3$ and $\omega=-\frac{1}{3}$. The solid, dashed and dot-dashed lines represent $\beta=0$, $0.04$ and $0.07$ respectively.}
\end{figure}

Causality condition states that the speed of sound must lies below the speed of light within the stellar interior. In the relativistic limit speed of light is taken as $c=1$. In case of anisotropic star, the causality conditions can be expressed as  $0\leq v_{r}^2(=\frac{dp_r^{total}}{d\rho^{total}})\leq1$ and $0\leq v_{t}^2(=\frac{dp_t^{total}}{d\rho^{total}})\leq1$, where $v_r$ and $v_t$ are the radial and transverse sound velocities, respectively. The causality conditions are shown graphically within the interior of the star $4U~1608-52$ in Figs.~\ref{fig19} and \ref{fig20} taking parametric choices of $m_s$, $B_g$ and $\beta$ for $\omega=-1$ and $-\frac{1}{3}$, respectively. From Figs.~\ref{fig19} and \ref{fig20}, we note that both sound speeds satisfy the causal limits. Interestingly, for $\beta=0$, $v_r^2=\frac{1}{3}$ throughout the system, as expected for a strange star, but with increase of dark energy, $v_r^2$ decreases, although it remains at a constant value from the centre to the surface. Additionally, with increase of $\beta$, $v_t^2$ at the centre decreases. The non-monotonic or counter intuitive behaviour of the sound speed arises from the combined effect of equation of state and the additional negative-pressure component. In our model, the presence of the dark-energy–like term (through $\beta$ and $\omega$) and the finite strange quark mass ($m_s\neq 0$) modifies the local relation between pressure and energy density. Near the centre, where the dark-energy contribution becomes comparatively significant, the effective stiffness of the EoS is reduced, leading to a lower sound speed than naively expected. Moving outward, the relative contribution of this negative-pressure component decreases, allowing the quark matter component to dominate and increase the effective sound speed locally. Thus, the behaviour reflects a contest between quark matter pressure and the additional negative-pressure component, rather than a simple monotonic dependence on density. All these variations of $v_r^2$ and $v_t^2$ are distinctively larger for $\omega=-1$ than for $\omega=-\frac{1}{3}$. Again, this trend in the sound velocity rules out the possibility of Laplacian instabilities in the model. 

\section{Stability analysis}
\label{sec6}
For a realistic model of a compact star, composed of a finite strange quark mass along with dark energy, the following stability criteria must hold good from the centre to the surface of the star:
\begin{enumerate}
	\item Tolman-Oppenheimer-Volkoff equation in a generalised form
	\item Herrera's cracking condition  
	\item Value of the adiabatic index and
	\item Stability against small radial oscillation.
	\item Calculation of Tidal Love Number 
\end{enumerate}

\subsection{Tolman-Oppenheimer-Volkoff equation}
\label{sec6.1}
\begin{figure}
	\centering
	\begin{minipage}{0.45\textwidth}
		\centering
		\includegraphics[width=0.85\linewidth,height=0.6\textwidth]{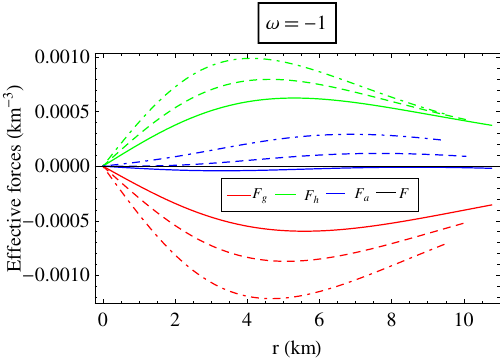}
		\subcaption{}\label{fig22}
	\end{minipage}\hfil
	\begin{minipage}{0.45\textwidth}
		\centering
		\includegraphics[width=0.85\linewidth,height=0.6\textwidth]{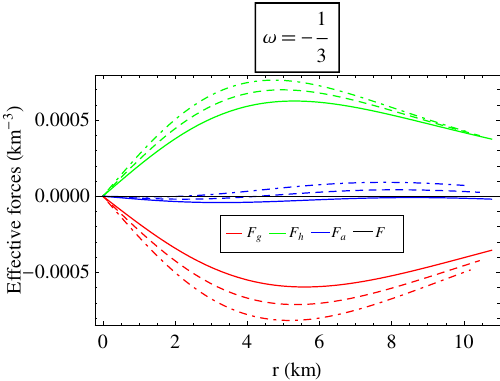}
		\subcaption{}\label{fig23}
	\end{minipage}\hfil
	\caption{Panel (a) represents variation of different forces with $r$ inside $4U~1820-30$ for $m_s=100~MeV$, $B_g=60~MeV/fm^3$, taking $\omega=-1$ and in Panel (b) for $m_s=100~MeV$, $B_g=60~MeV/fm^3$, taking $\omega=-\frac{1}{3}$. The solid, dashed and dotdashed lines represent values of $\beta$ as, $0.0$, $0.04$ and $0.07$ respectively. Here, the red, blue and green lines represent $F_g$, $F_a$ and $F_h$, respectively.}
\end{figure}

The TOV equation \cite{Tolman,Oppenheimer} in the presence of dark energy can be represented in general as: 
\begin{equation}
	-\frac{M_{G}(r)(\rho^{total}+p_r^{total})}{r^2}e^{(\lambda-\nu)/2}-\frac{dp_r^{total}}{dr}+\frac{2}{r}(p_t^{total}-p_r^{total})=0,\label{42}
\end{equation}
where $\lambda$ and $\nu$ are the metric potentials given in eqs.~(\ref{29}) and (\ref{32}), respectively. Following the formula of Tolman-Whittaker \cite{ogron} and using the EFE, $M_{G}(r)$, the effective gravitational mass, encompassed within the spherical region having radius $r$, is given as follows: 
\begin{equation}
	M_{G}(r)=\frac{1}{2}r^{2}\nu^{\prime}e^{(\nu-\lambda)/2}.\label{43}
\end{equation}
Using eq.~(\ref{43}), eq.~(\ref{42}), can be reduced to:
\begin{equation}
	-\frac{\nu^{\prime}}{2}\left(\rho^{total}+p_r^{total}\right)-\frac{dp_r^{total}}{dr}+\frac{2}{r}\left(p_t^{total}-p_r^{total}\right)=0.\label{44}
\end{equation}
The modified Tolman-Oppenheimer-Volkoff equation in the presence of dark energy as given in eq.~(\ref{44}), represents the equilibrium condition of fluid sphere under the combined effect of forces present inside a star, namely, gravity force, denoted by $F_{g}$, hydrostatic force, denoted by $F_{h}$ and the anisotropic force, denoted by $F_{a}$, where $F_g=-\frac{\nu^{\prime}}{2}(\rho^{total}+p_r^{total})$, $F_h=-\frac{dp_r^{total}}{dr}$, $F_a=\frac{2\Delta}{r}$ with $\Delta=p_t^{total}-p_r^{total}$ is the total pressure anisotropy. It follows from eq.~(\ref{45}) that inside the star, 
\begin{equation}
	F_g+F_h+F_a=0.\label{45}
\end{equation}
The radial gradient of pressure ($-\frac{dp_r^{total}}{dr}$) represents a force per unit volume $(-\nabla p^{total})$ arising from spatial imbalance of internal stresses in a fluid. In compact stars, this pressure-gradient force balances gravity (determined by the energy density), as formalized by the Tolman–Oppenheimer– Volkoff equation derived from energy–momentum conservation. Anisotropic force, is defined as $\frac{2\Delta}{r}$, where, $\Delta$ is the difference between transverse and radial pressure. So, the term  $\frac{2\Delta}{r}$ indicates pressure gradient and thus represents a force. These quantities do not have the physical units of force. That is why we have transformed the unit in terms of $km^{-3}$. In Figs.~\ref{fig22} and \ref{fig23}, we have plotted the variation of these forces by taking suitable parametric choices of $m_s$, $B_g$ and $\beta$ for $\omega=-1$ and $\omega=-\frac{1}{3}$, respectively. It is noted that condition of static equilibrium is satisfied under the combined effect of the all forces inside the stellar configuration. It is observed that gravity force ($F_g$) is balanced by the resultant effect of the hydrostatic force ($F_h$) and the anisotropic force ($F_a$) in presence of non zero $m_s$ and dark energy. So we may conclude that TOV equation holds good in the model also. 

\subsection{Herrera cracking condition}
\label{sec6.2}
Fluid configuration of anisotropic nature inside a compact object should follow the stable equilibrium condition in presence of small fluctuations in the physical variables associated with fluid. The stability criterion of a fluid of anisotropic nature may be explained using the cracking concept of a self gravitating object proposed by Hererra \cite{Herrera2}. Abreu \cite{Abreu} derived a criterion based on Hererra's concept which states that a fluid configuration will be in the condition of stable equilibrium if the square of radial ($v_{r}^{2}$) and tangential ($v_{t}^{2}$) sound speeds fulfil the following condition:
\begin{equation}
	0 \leq |v_{t}^{2}-v_{r}^2| \leq 1. \label{46}
\end{equation}
\begin{figure}
	\centering
	\includegraphics{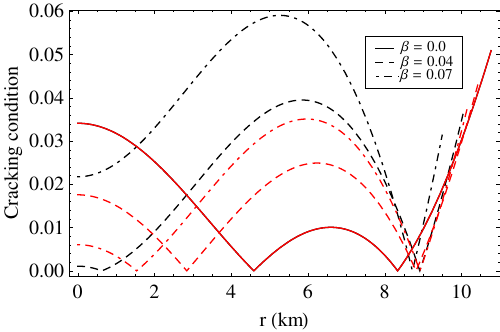}
	\caption{Variation of $|v_t^2-v_r^2|$ with $r$ inside $4U~1608-52$ for different values of $\beta$ using a parametric choice of $m_s=100~MeV$ and $B_g=60~MeV/fm^3$. Here, the solid, dashed and dot-dashed lines indicate values of $\beta=0.0,~0.04$ and $0.07$, respectively. The black and red lines represent $\omega=-1$ and $\omega=-\frac{1}{3}$, respectively.}\label{fig24}
\end{figure}

In Fig.~\ref{fig24} the Abreu criterion is shown for the star $4U~1608-52$ and it is evident that the condition given in Eq.~(\ref{46}) is satisfied for both $\omega=-1$ and $\omega=-\frac{1}{3}$. Thus, we may conclude that the stellar configuration remains in stable equilibrium in presence of dark energy as well as non-zero strange quark mass ($m_s\neq0$).

\subsection{Adiabatic index}
\label{sec6.3}
In case of of a star composed of three flavour quark matter and dark energy the adiabatic index $\Gamma$ is defined as follows: 
\begin{equation}
	\Gamma=\frac{(\rho^{total}+p_r^{total})}{p_r^{total}}(\frac{dp_r^{total}}{d\rho^{total}})=\frac{(\rho^{total}+p_r^{total})}{p_r^{total}}v_r^{2}. \label{47}
\end{equation}
For stable isotropic fluids, Heintzmann and Hillebrandt \cite{HH} established that $\Gamma>\frac{4}{3}$ (Newtonian limit). on the other hand, according to the work of Chan et al. \cite{CHEN}, the value of $\Gamma$ picks up higher value for a relativistic anisotropic fluid given by $\Gamma>\Gamma^{\prime}_{max}$, where,
\begin{equation}
	\Gamma^{\prime}_{max}=\frac{4}{3}-\left[\frac{4}{3}\frac{(p_r^{total}-p_t^{total})}{|(p_r^{total})^{\prime}|r}\right]_{max}. \label{48}
\end{equation}
\begin{figure}
	\centering
	\includegraphics{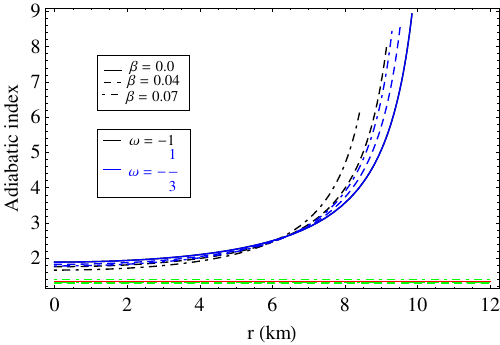}
	\caption{Radial variation of the adiabatic index ($\Gamma$) inside $4U~1608-52$ for different values of $\beta$ with $B_g=60~MeV/fm^3$, $m_s=100~MeV$. Here, the solid, dashed and dotdashed lines represent the radial variation of $\Gamma$ for $\beta=0.0, 0.04$ and $0.07$, respectively. The solid red and green lines indicate the values of $\Gamma=\frac{4}{3}$ and $\Gamma^{\prime}_{max}$, respectively. Also the black and blue lines represent the variation of $\Gamma$ for $\omega=-1$ and $-\frac{1}{3}$, respectively.}\label{fig25}
\end{figure}

In Fig.~\ref{fig25}, the variation of the adiabatic index ($\Gamma$) with $r$ is shown, and we note that the requirement $\Gamma>\Gamma^{\prime}_{max}$ holds good inside the star $4U~1608-52$ even if dark energy is present. Therefore, our model is stable in terms of the adiabatic index.

\subsection{Study of stability against small radial oscillation:}
\label{sec6.4}
The study of Lagrangian perturbation in radial pressure at the stellar surface ($r=R$) is an another convenient way to check the stability of a stellar configuration. On this approach radial pressure is perturbed and the frequencies of the normal mode of oscillation ($\omega_{0}$) is evaluated. According to the work of Pretel \cite{pretel} and assuming that the changes are adiabatic, the coupled equations responsible for oscillations of infinitesimal radial mode are governed by the following equations:
\begin{equation}
	\frac{d\zeta(r)}{dr}=-\frac{1}{r}\left(3\zeta(r)+\frac{{\Delta}p_r^{total}}{{\gamma}p_r^{total}}\right)+\frac{1}{2}\frac{d{\nu}}{dr}\zeta(r),\label{49}
\end{equation}
\begin{eqnarray}
	\frac{d({\Delta}p_r^{total})}{dr}=\zeta\Bigg[\frac{\omega_0^2}{c^2}e^{(\lambda-\nu)}(\rho^{total}+p_r^{total})r-4\frac{dp_r^{total}}{dr} -k(\rho^{total}+p_r^{total})e^{\lambda}rp_r^{total}+\nonumber\\ \frac{r}{4}(\rho^{total}+p_r^{total})\left(\frac{d\nu}{dr}\right)^2\Bigg]
	-{\Delta}p_r^{total}\left[\frac{1}{2}\frac{d\nu}{dr}+\frac{k}{2}(\rho^{total}+p_r^{total})re^{\lambda}\right].\label{50}
\end{eqnarray}
where $k=\frac{8{\pi}G}{c^4}=1$ and $\left|{\Delta}p_r^{total}\right|$ represents the change in absolute value of the Lagrangian perturbation in radial pressure and the form of eigenfunction $\zeta(r)$ is expressed as $\zeta(r)=\frac{\xi(r)}{r}$, where $\xi(r)$ is termed as Lagrangian displacement. However, owing to spherical symmetry, $\xi(r)$ must vanish at $r=0$, i.e., $\xi(0)=0$. Following the work of Pretel \cite{pretel}, we adopt that $\zeta(r)$ is normalised, i.e., $\zeta(0)=1$. Furthermore, eq.~(\ref{49}) poses a singularity at $r=0$. Therefore, to solve eqs.~(\ref{49}) and (\ref{50}), the coefficient of $(\frac{1}{r})$ in eq.~(\ref{49}) must vanish as $r\rightarrow0$. This implies,
\begin{equation}
	{\Delta}p_r^{total}=-3{\gamma}{\zeta}p_r^{total}.\label{51}
\end{equation}
As pressure itself vanishes at the stellar surface ($r=R$), the Lagrangian perturbation in the radial pressure should also vanish, which implies 
\begin{equation} 
	\left|{\Delta}p_r^{total}(R)\right|=0.\label{52}
\end{equation}

\begin{figure}
	\centering
	\includegraphics{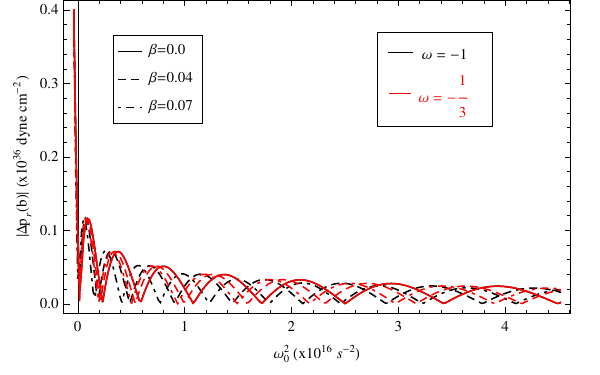}
	\caption{Plots of $\left|{\Delta}p_r(R)\right|$ with $\omega_0^2$ for $m_s=100~MeV$ and $B_g=60~MeV/fm^3$. Here, the solid, dashed and dotdashed lines indicate variations for $\beta=0.0,~0.04$ and $0.07$, respectively, inside $4U~1608-52$. The black and red lines represent the respective variation for $\omega=-1$ and $\-\frac{1}{3}$, respectively.}\label{fig26}
\end{figure} 
In Fig.~\ref{fig26}, the plot of $\left|{\Delta}p_r^{total}\right|$ vs. $\omega_0^2$ are shown for dark energy EoS parameter $\omega=-1$ and $\omega=-\frac{1}{3}$. Minima of these plots indicate the exact value of frequencies of normal mode. It is observed that all $\omega_0^2$ are positive, which indicates that the present model is stable under small radial oscillations in presence of dark energy as well as non-zero mass of strange quark.

\subsection{Tidal Love Number and tidal deformability}
\label{6.5}
The structural deformability of a self gravitating body like a NS, under the influence of an external tidal field is represented by the tidal deformability. When a compact object is in a binary system, the gravitational pull from its companion induces tidal deformation. The response of the star to these deformations is characterized by the Tidal Love Number (TLN). When a compact object is subjected to an external tidal field $\eta_{ij}$, it acquires an induced quadrupole moment $Q_{ij}$. The tidal deformability is defined through the relation $\lambda_{tidal-def}=-\frac{Q_{ij}}{\eta_{ij}}$. The dimensionless tidal deformability ($\Lambda$) is then expressed as $\Lambda=\frac{\lambda_{tidal-def}}{M^5}$, $M$ being the mass contained within radius $R$ of the compact star. Moreover, $\Lambda$ is related to $l=2$ dimensionless Tidal Love Number $k_2$ as \cite{THIN,KC}:
\begin{equation}
	k_2=\frac{3}{2}\Lambda\Big(\frac{M}{R}\Big)^5.\label{53}
\end{equation}
Now, considering linear perturbations of the background space-time metric, following the approach of Thorne and Campolattaro \cite{KST}, the modified metric in presence of linearized small perturbation ($h_{ij}$) can be represented as:
\begin{equation}
	\tilde{g}_{ij}=g_{ij}+h_{ij}. \label{54}
\end{equation}
Adopting the formalism developed by Regge and Wheeler \cite{TR} and later applied by Biswas and Bose \cite{BB}, we focus on static, even-parity perturbations with $l=2$ and $m=0$. Following these articles, the perturbation ($h_{ij}$) can be written as:
\begin{equation}
	h_{ij}=diag[H_0e^\nu,H_2(r)e^\lambda,r^2K(r), r^2\sin2\theta\;\;K(r)]Y_{2m}(\theta,\phi).\label{55}
\end{equation} 
Now, following the procedure as given in the Refs.~\cite{BB,TH}, the basic equation for $H(r)$ in presence of anisotropy can be written as:
\begin{equation}
	H^{\prime\prime}(r)+f_1 H^{\prime}(r)+f_2 H(r)=0,\label{56}
\end{equation}
where, 
\begin{equation}
	f_1=\frac{2}{r}+e^{\lambda}\Big(\frac{2m(r)}{r^2}+0.5r(p_r^{total}-\rho^{total})\Big),\label{57}
\end{equation}
and
\begin{equation}
	f_2=-\frac{6e^{\lambda}}{r^2}+0.5e^{\lambda}\Bigg(4\rho^{total}+8p_r^{total}+\frac{\rho^{total}+p_r^{total}}{A\;v_r^2}\left(1+v_r^2\right)\Bigg)-\Big(\frac{d\nu}{dr}\Big)^2.\label{58}
\end{equation} 
where, $A=\frac{dp_t^{total}}{dp_r^{total}}$ and $v_r^2$ is the squared radial sound velocity. Outside the star, the metric is the Schwarzschild one and hence by setting the parameters $p_r^{total}=0$, $p_t^{total}=0$, $\rho^{total}=0$ and $e^{-\lambda}=(1-\frac{2M}{R})$, the basic eq.~(\ref{56}) reduces to the following form, 
\begin{equation}
	H^{\prime\prime}+\frac{2(M-r)H^{\prime}(r)}{r(2M-r)}-\frac{2H(r)\left(2M^2-6Mr+3r^2\right)}{r^2\left(r-2M\right)^2}=0,\label{59}
\end{equation}
The solution of eq.~(\ref{59}) can be put in the form,
\begin{eqnarray}
	H(r)=c_1\left(\frac{r}{M}\right)^2\left(1-\frac{2M}{r}\right)\Bigg[-\frac{M(M-r)(2M^2+6Mr-3r^2)}{r^2(2M-r)^2}+\frac{3}{2}\log\Big(\frac{r}{r-2M}\Big)\Bigg]+3c_2\left(\frac{r}{M}\right)^2\left(1-\frac{2M}{r}\right).\label{60}
\end{eqnarray}
In order to evaluate the constants $c_1$ and $c_2$ we now make a series expansion of eq.~(\ref{60}) so that at very large distance $(r)$, the asymptotic behaviour of eq.~(\ref{60}) becomes,
\begin{equation}
	H(r)=\frac{8}{5}c_1\left(\frac{M}{r}\right)^3\mathcal{O}\left(\frac{r}{M}\right)+3c_2\left(\frac{r}{M}\right)^2+\mathcal{O}\Bigg(\left(\frac{M}{r}\right)^4\Bigg).\label{61}
\end{equation}
Now at large $r$ in the star's asymptotic rest frame the $g_{tt}$ metric coefficient is given by \cite{KST,WMS,TH}
\begin{equation}
	\frac{1-g_{tt}}{2}=-\frac{M}{r}-\frac{3Q_{ij}}{2r^3}\left(n^in^j-\frac{\delta^{ij}}{3}\right)+\mathcal{O}\left(\frac{1}{r^3}\right)+\frac{1}{2}\eta_{ij}x^ix^j+\mathcal{O}(r^3),\label{62}	
\end{equation}
where, $n^i=\frac{x^i}{r}$. The unknowns $c_1$ and $c_2$ can be determined from the matching of the asymptotic solution in eq.~(\ref{61}) together with eq.~(\ref{62})and the relation $Q_{ij}=-\lambda_{tidal-def}\eta_{ij}$. Thus we get, 
\begin{equation}
	c_1=\frac{15Q}{8M^3}, \;\;\;\;\;\;\;\;\;\;c_2=\frac{M^2\eta}{3}.\label{63}
\end{equation}
Now the expression of the Tidal Love Number ($k_2$) can be obtained from eqs.~(\ref{53}), (\ref{61}) and (\ref{63}) at the surface of a compact object as 
\begin{eqnarray}
	k_2=\frac{8u^5}{5}\left(1-2u\right)^2\Bigg(2+2u(y-1)-y\Bigg)\Bigg[2u\{6-3y+3u(5y-8)\}+4u^3\Biggl\{13-11y\nonumber\\+u(3y-2)+2u^2(1+y)\Biggr\} +3(1-2u)^2\Biggl\{2-y+2u(y-1)\log(1-2u)\Biggr\}\Bigg]^{-1}.\label{64}
\end{eqnarray}
where, $y=\frac{R H^{\prime}(R)}{H(R)}$, $u=\frac{M}{R}$ and $H^{\prime}(R)$ being the value of $H(r)$ at the stellar boundary ($r=R$). From the binary neutron star merger event $GW170817$, Abbott and Abbott \cite{BPA2} derived observational limits on the dimensionless tidal deformability parameter ($\Lambda$) providing crucial constraints on the EoS of compact star. Furthermore, Bauswein et al. \cite{ABA} reported that for a compact star of canonical mass $1.4~M_\odot$ the tidal deformability satisfies the upper bound, $\Lambda_{1.4}<800$. Following the recent LIGO-Virgo-KAGRA (LVK) data driven approach presented by Ref.~\cite{CH}, it has been shown that the analysis of GW170817 posterior may provide an EoS independent constraint of tidal deformability given as $\Lambda_{1.4}=222.89^{+420.33}_{-98.85}$. 
Further, combining these results with x-ray EoS independent bounds, a multi-messenger limit of $\Lambda_{1.4}=265.18^{+237.88}_{-104.38}$ is obtained. Tidal deformability is a measure of stability of compact star and accordingly, we have tried to investigate the variation of tidal deformability with dark energy component.\par 

In Table~\ref{tab6}, we present the computed values of the Tidal Love Number ($k_2$) and the corresponding $\Lambda$ for different compact star configurations. Moreover, in the high density regime inside compact stars a phase transition is anticipated. However, if multiple charges such as electric charge, baryon number density, number density of quarks remain conserved during a phase transition, the global charge neutrality relation may result in a continuous variation of pressure with energy density in case of first order phase transition too. Further, from the inner surface to the outer surface of the star a discontinuous density distribution may prevail. In this regard Damour and Nagar \cite{TD} showed that such density discontinuity greatly affects the Tidal Love number. In light of the above, we have reformulated the tidal parameters by considering a finite surface density correction. To start with, we express the modified sound velocity in the following form,
\begin{equation}
	\frac{d\rho^{total}}{dp_r^{total}}=\frac{1}{v_r^2}=\frac{d\rho^{total}}{dp_r^{total}}\Bigg{|}_{p_r^{total}\neq (p_r^{total})d}+\Delta \rho \delta (p_r^{total}-(p_r^{total})_d).\label{65}
\end{equation}
where, $(p_r^{total})_d=$ pressure at the discontinuity, $\delta \rho =\rho^{total}((p_r^{total})_d+0)-\rho^{total}((p_r^{total})_d-0)=$ discontinuity in energy density. Considering appropriate boundary conditions, we obtain the surface density corrected $y$ in the following form \cite{SP2} 
\begin{equation}
	y(R+\epsilon)=y(R-\epsilon)-\frac{\rho^{total}(R+\epsilon)-\rho^{total}(R-\epsilon)}{m(R)/4\pi R^3}.\label{66}
\end{equation}

Using eq.~(\ref{66}) we have obtained the surface density corrected tidal parameters and are tabulated in Table~(\ref{tab6}).
\begin{figure}
	\centering
	\includegraphics{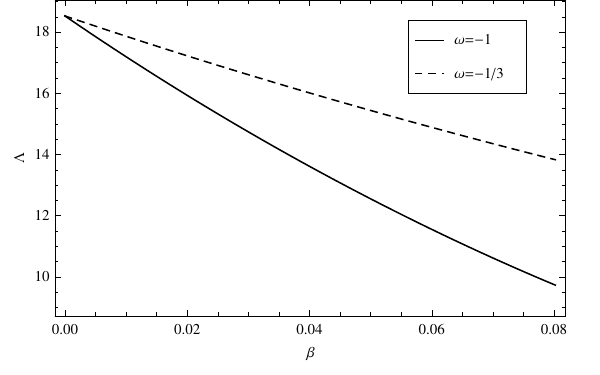}
	\caption{Variation of tidal deformability ($\Lambda$) with dark energy parameter ($\beta$) for the compact object $4U~1608-52$ with a parametric choice of $B_g=60~MeV/fm^3$ and $m_s=100~MeV$. The solid and dashed lines represent the variation for $\omega=-1$ and $-\frac{1}{3}$, respectively. }\label{fig27}
\end{figure}

\begin{table}[ht]
	\centering
	\caption{Table for tidal deformability of different compact objects in presence of dark energy for finite strange quark mass ($m_s=100~MeV$) and $B_g=60~MeV/fm^3$.}\label{tab6}
	\resizebox{0.8\textwidth}{!}{$
		\begin{tabular}{@{}c|c|cccc|cccc}
			\hline
			Name of Compact  & Observed   & \multicolumn{4}{c}{$\omega=-1$} \vline & \multicolumn{4}{c}{$\omega=-\frac{1}{3}$}\\ \cline{3-10}
			stars            &  mass            & $\beta$ & $k_2$     & $\Lambda$ & $\Lambda_{corr}$   & $\beta$ & $k_2$   & $\Lambda$ & $\Lambda_{corr}$\\
			&  $(M_\odot)$     &         &           &              &         &         &           \\ \hline
			$Vela~X-1$       & 1.77\cite{TG}    & 0.068   & 0.0223    & 9.821 &  22.437   & 0.0801  & 0.0217    & 12.673 & 36.295\\
			$4U~1608-52$     & 1.74\cite{Guver} & 0.080   & 0.0233    & 9.745 &  20.753  & 0.080  & 0.0221     & 13.836 & 40.443\\
			$Cen~X-3$        & 1.49\cite{TG}    & 0.072   & 0.0270    & 22.831 & 60.278  & 0.0800  & 0.0248     & 28.697 & 100.441\\
			$HER~X-1$        & 0.85\cite{TG}    & 0.060   & 0.0278    & 218.258 & 1137.64 & 0.0800  & 0.0258    & 243.131 & 1580.49\\ 
			$PSR~J-1903+327$ & 1.667\cite{TG}   & 0.069   & 0.0242    & 13.456 & 32.450  & 0.0801  & 0.0230     & 17.102  & 52.536\\ 
			$4U~1820-30$     & 1.58\cite{TG}    & 0.078   & 0.0262    & 16.518 & 39.226  & 0.0801  & 0.0239     & 22.031  & 72.029\\  
			$HESS-1731-347$  & 0.77\cite{PC}    & 0.050   & 0.0262    & 320.484 & 2043.46 & 0.0600  & 0.0244    & 355.052 & 2814\\ 
			$PSR~J0030+451$  & 1.34\cite{TER}   & 0.055   & 0.0270    & 40.546 & 138.081  & 0.0680  & 0.0282    & 46.541  & 193.227 \\ \hline   
		\end{tabular}$}	
\end{table}

From Table~\ref{tab6} we note that the Tidal parameters obtained in this model are in well agreement with the estimated bounds obtained from the binary coalescences. Again, to study the effect of $\beta$ on $\Lambda$, we have graphically represented the variation of $\Lambda$ with $\beta$ for a parametric choice of $B_g$ and $m_s$ as shown in Fig.~\ref{fig27}. From Fig.~\ref{fig27}, it is noted that increasing $\beta$ decreases the tidal deformability ($\Lambda$). The decrease in $\Lambda$ with increasing dark energy fraction indicates that the star become more compact and less deformable, indicating that dark energy contributes to tightening the structure of the compact star, possibly through gravitational self binding or a modified pressure balance. Again a stiffer EoS for dark energy ($\omega=-1$) indicate more compact and less deformed structure.  

\section{Disscusion}
\label{sec7}
In this paper, we have studied the hybrid star composed of dark energy and three flavour quark matter with non-zero strange quark mass ($m_s\neq0$). Considering the assembly of quarks as a Fermi gas, which is overall charge neutral, the EoS for quark matter is taken as proposed in the MIT bag model \cite{CHK,AC}. However, owing to the presence of a finite $m_s$ value, the MIT bag EoS is modified and given in Eq.~(\ref{19}). Again, choosing the EoS of the dark energy, as given in eq.~(\ref{27}), the effective EoS for the system composed of dark energy and three flavour quarks is given in eq.~(\ref{28}). To solve the EFE, we choose the $g_{rr}$ component of the metric as given by Finch-Skea (eq.~\ref{29}). Using this value of $g_{rr}$ and the composite EoS of the system given in eq.~(\ref{28}), we have derived the various physical parameters of the star such as, total energy density, total radial as well as transverse pressure and total anisotropy. Now, to maintain the overall causality criterion, the fraction of dark energy ($\beta$) present within the star must lie within the range $0<\beta\leq-\frac{1}{3\omega}$, which yields $\beta\leq 1$ for $\omega=-\frac{1}{3}$ and $\beta\leq{0.33}$ for $\omega=-1$. This implies that the maximum percentage of dark energy ($\frac{\beta}{1+\beta}\times100$) that can be attributed is 50 $\%$ if $\omega=-\frac{1}{3}$ and $25$ $\%$ for $\omega=-1$. To examine the stability, energy per baryon ($E_B$) at zero pressure condition is evaluated for different parametric choices of bag values in presence of dark energy and $m_s=100~MeV$. The variation of $E_B$ with $\beta$ is shown in Figs.~\ref{fig1} and \ref{fig2}, respectively, for $\omega=-1$ and $-\frac{1}{3}$. Based on the $E_B$ behaviour, the permissible range of $\beta$ is further restricted under the constraint that $E_B\leq 930.4~MeV$ for stable and $930.4<E_B\leq 939~MeV$ for metastability. A summary of these findings is presented in Table~\ref{tab1}. To find the maximum mass, we have solved the TOV equations using the effective EoS for our model. The mass-radius plot is shown in Fig.~\ref{fig3}. In Table~\ref{tab2} maximum mass ($M_{max}$), radius ($R_{max}$) and central density are tabulated taking $B_g=60~MeV/fm^3$ and different parametric choices of $m_s$ and $\beta$. It is observed that with the increase of $m_s$ and $\beta$, maximum mass and radius both decreases indicating the softer nature of the EoS. From Table~\ref{tab2}, it is evident that for a fixed $m_s$, increasing the dark energy coupling parameter $\beta$ results in a decrease of maximum mass and radius. This can be attributed to the fact that with increasing $\beta$, the percentage of quark energy converting to dark energy increases. Therefore, the matter content contributing to the maximum mass decreases which refers to a softer EoS validating the decrease in maximum mass and radius. The stability criterion as proposed by Harrison-Zeldovich-Novikov states that mass of a star should increase with the increase of central density, i.e., ($\frac{dM}{d\rho_c^{total}})>0$ \cite{YBZ,BKH}. From Fig.~\ref{fig4}, it is noted that the stability criterion ($\frac{dM}{d\rho_c^{total}})>0$ holds up to a local maxima. Beyond this point, ($\frac{dM}{d\rho_c^{total}})<0$ showing the instability in the fluid configuration. In Fig.~\ref{fig4a}, the variation of maximum mass with possible range of bag parameter is shown. Figs.~\ref{fig5} and \ref{fig6} show the variation of $M_{max}$ and $R_{max}$ with the percentage of dark energy present for a parametric choice of $m_s=100~MeV$, $120~MeV$ and $B_g=60~MeV/fm^3$ for both $\omega=-1$ and $\omega=-\frac{1}{3}$. Additionally, it is clear that $M_{max}$ is relatively smaller for $\omega=-1$ than for $\omega=-\frac{1}{3}$, for a fixed value of $m_s$ and $\beta$. Therefore, to account for higher value of the maximum mass, a stiffer the dark energy EoS is preferable, when $m_s$ is fixed at some value. The maximum mass attainable in this model is $2.012~M_{\odot}$ for $m_s=0~MeV$, $\beta=0$, $B_g=57.55~MeV/fm^3$ and $\omega=-\frac{1}{3}$. Thus the mass of the massive pulsar $PSR~J0348+0432$ \cite{JAN} may be accommodated in this model for suitable choice of model parameters. \par 

To show the effects of $m_s$ and $\beta$ on the metric potentials, we have graphically represent $e^{-\lambda}$ and $e^{\nu}$ in Figs.~\ref{fig7} and \ref{fig8} for the selected source $4U~1608-52$ for $\omega=-1$ and $\omega=-\frac{1}{3}$, respectively by taking a suitable parametric set of value for $\beta$, $m_s$ and $B_g$. Therefore, it can be concluded that a stiffer or softer dark energy EoS has some effects on the total mass and radius of a star. To observe the effect of dark energy in the presence of non-zero $m_s$ value, we have graphically plotted both the quark part and the total energy density and radial pressure for the selected star $4U~1608-52$ in Figs.~\ref{fig9}-\ref{fig13} and it is evident that density and radial pressure are monotonically decreasing from centre to surface, having a maximum value at the centre. The radial variation of total tangential pressure is depicted in Fig.~\ref{fig14}, which exhibits that near the centre, $p_t^{total}$ increases with increasing $\beta$. The total anisotropy profile, shown in Fig.~\ref{fig15} reveals that at the centre, $\Delta$ is zero, satisfying the stable configuration criterion. As $\rho^{total}$ and compactness ($u$) both increase with $\beta$, hence the radius of a star for a given mass ($M$) must decrease as indicated in Fig.~\ref{fig16}. In Table~\ref{tab3}, the predicted radii of different stars are tabulated and we note that estimated radii from recent observations may also be predicted from present model with suitable choices of parameters $\beta$ and $\omega$. Such results show that a correlation exists between $\beta$, $m_s$ and $B_g$. It is observed from Table~\ref{tab3}, that estimated radius from observation may predicted pretty well from the model for $\omega=-1$ rather than $\omega=-\frac{1}{3}$. Therefore, comparing the radii estimated from recent observations and predicted from the present model as tabulated in Table~\ref{tab3}, it may be concluded that such stars may contain a decent amount of dark energy as well as strange quark of non-zero mass. The variation of the predicted radius for $4U~1608-52$ with dark energy percentage is shown in Fig.~\ref{fig16} for different suitable parametric combinations of $m_s$ and $B_g$. The energy conditions inside $4U~1608-52$ are shown graphically in Figs.~\ref{fig17} and \ref{fig18} considering a parametric choice of $m_s=100~MeV$ and $B_g=60~MeV/fm^3$ and $\beta$ as a free parameter for the respective cases of $\omega=-1$ and $-\frac{1}{3}$. It is evident that in the presence of dark energy and non-zero $m_s$, all the energy conditions are satisfied from the centre to the surface. Different physical quantities, such as the total central energy density ($\rho_c^{total}$), total central radial pressure ($p_{rc}^{total}$), mass to radius ratio ($u$) and surface redshift ($Z_s$), for the selected star candidates are listed in Table~\ref{tab4} which are assumed to be hybrid star. From Table~\ref{tab4}, it is clear that $u$ and $Z_s$ lie within the allowed range,as required for a stable configuration.\par 
\begin{figure}
	\centering
	\includegraphics{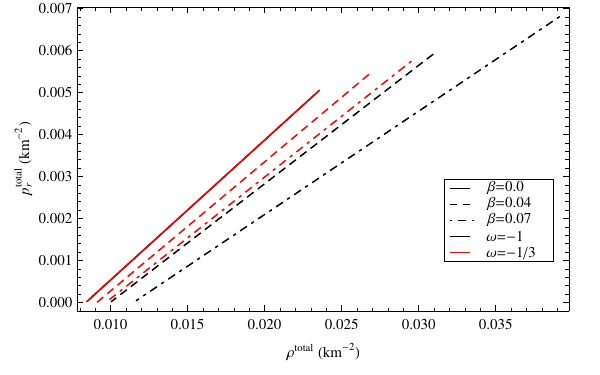}
	\caption{EoS plot of $4U~1608-52$ for a parametric choice of $B_g=60~MeV/fm^3$, $m_s=100~MeV$ and $\omega=-1$ (black line) and $\omega=-\frac{1}{3}$ (red line). The solid, dashed and dotdashed lines represent $\beta=0.0,~0.04$ and $0.07$, respectively.}\label{fig28}
\end{figure}

The radial variation of $v_r^2$ and $v_t^2$ are given in Figs.~\ref{fig19} and \ref{fig20}, for $\omega=-1$ and $\omega=-\frac{1}{3}$, respectively. It is evident that $v_r^2=\frac{1}{3}$ throughout the star for $\beta=0$ in both the cases of $\omega=-1$ and $\omega=-\frac{1}{3}$, as expected for the quark star configuration \cite{PBED}. However, with increasing $\beta$, $v_r^2$ decreases although this modified value remains constant throughout the star. It is also shown that $v_t^2$ picks up different values as $\beta$ changes. The relative changes in the sound velocity are greater for $\omega=-1$ than for $\omega=-\frac{1}{3}$, as evident form Figs.~\ref{fig19} and \ref{fig20}. The stability of the model consisting of non-zero $m_s$ admixed with dark energy is studied through different context such as: TOV equation in hydrostatic equilibrium, cracking condition proposed by Herrera \cite{Herrera2} and adiabatic index for both the cases $\omega=-1$ and $-\frac{1}{3}$. From Figs.~\ref{fig22} and \ref{fig23}, it is evident that the TOV equation as given in Eq.~(\ref{45}) holds good in this model in presence of $m_s$ and dark energy. The cracking condition for $4U~1608-52$ is also verified and shown in Figs.~\ref{fig24}. The value of adiabatic index is checked for suitable choices of $\beta$ and is shown in Fig.~\ref{fig25} and it is evident that the model is stable and physically acceptable. In Fig.~\ref{fig26}, we have also studied the Lagrangian change in radial pressure at the surface of $4U~1608-52-30$ against $\omega_0^2$ for the parametric choice of $B_g$, $m_s$ and different $\beta$ value. The minima of these plots correspond to the exact eigen frequencies of different normal modes. Notably, the frequency spectrum is real ($\omega_0^2>0$), indicating a stable configuration. Now, to further extent of stability study, tidal deformability has been considered. In Table~\ref{tab6}, we have tabulated the Tidal Love Number ($k_2$) and tidal deformability ($\Lambda$) of the selected stars. We observe that the tidal parameters obtained in this study, as charted in Table\ref{tab6}, satisfy the LVK and multi-messenger constraints. Also, effect of $\beta$ on $\Lambda$ is studied for the chosen star $4U~1608-52$, as shown in Fig.~\ref{fig27} for parametric choice of model parameters. The tidal deformability of star having canonical mass $1.4~M_{\odot}$ is evaluated in this model as $74.55$ which falls within the range, $70<\Lambda_{1.4}<580$ \cite{BPA3} considering the gravitational wave constraints. Now, the simultaneous increase in compactness of the selected star, which is evident from Fig.~\ref{fig16} and decrease in deformation ($\Lambda$) with increasing dark energy as given in Fig.~\ref{fig27}, suggests that dark energy contributes to make the star more compact, less deformed and stable. Finally, in Fig.~\ref{fig28}, we have plotted the variation of the total radial pressure against the total energy density for the selected source $4U~1608-52$ for suitable parametric choices of $B_g$, $m_s$ and different $\beta$ value with $\omega=-1$ and $-\frac{1}{3}$, respectively. From Fig.~\ref{fig28}, it is evident that for a fixed value of $\omega$, the composite EoS for the system becomes softer for an increase in $\beta$. Again, for a particular value of $\beta$, the EoS becomes stiffer for $\omega=-\frac{1}{3}$, in comparison to $\omega=-1$.\par 

The inclusion of non-zero value of $m_s$ introduces an intrinsic asymmetry between light and strange quark contributions, which becomes particularly relevant at intermediate densities where the strange quark population is gradually activated. The combined effect of $m_s$, the dark energy coupling parameter $\beta$, and the dark energy equation of state parameter $\omega$ leads to non-trivial modifications in the stiffness of the equation of state. Specifically, a finite $m_s$ tends to soften the EoS at lower densities due to the mass threshold for strange quark production, while at higher densities, its contribution enhances the overall energy density, thereby influencing the stability window. When coupled with dark energy (through $\beta$) and its negative pressure contribution (governed by $\omega$), this can shift the critical central density at which stable configurations transition to unstable ones. This interplay can be interpreted as an effective competition: $m_s$ delays the onset of strange quark dominance, whereas dark energy tends to counteract gravitational collapse. As a result, the stability region in the parameter space ($m_s$,$\beta$ and $\omega$) is modified, potentially allowing stable configurations at higher central densities than in the massless case. This also has implications for phase structure, as the presence of finite $m_s$ may smoothen or shift the transition between two-flavour and three-flavour quark matter, which in turn is influenced by the strength of dark energy coupling.\par 
Our primary objective is to explore the possible implications of incorporating dark energy effects, together with a finite strange quark mass $(m_s)$, within compact star modeling. In this sense, the current study should be viewed as a phenomenological investigation rather than a definitive predictive model. The inclusion of dark energy (via $\beta$ and $\omega$) in combination with $m_s\neq0$ does introduce potentially distinguishable trends. For instance, the model tends to allow relatively higher compactness compared to standard quark star models without dark energy. Additionally, the interplay between $m_s$, $\beta$, and $\omega$ can lead to shifts in the stability window and subtle changes in the mass–radius slope, which could, in principle, be tested against more precise future observations.

\section*{Acknowledgment}
RR is grateful to the Department of Physics, Coochbehar Panchanan Barma University, for providing necessary help to carry out the research work. DB is grateful to the Department of Science and Technology (DST), Govt. of India, for providing the fellowship vide no: DST/INSPIRE/Fellowship/2021/IF210761. PKC gratefully acknowledges support from IUCAA, Pune, India, under the Visiting Associateship Programme. We are thankful to the annoyamous referees for the valueable comments. This research was funded by Princess Nourah bint Abdulrahman University Researchers Supporting Project number (PNURSP2026R443), Princess Nourah bint Abdulrahman University, Riyadh, Saudi Arabia.

\end{document}